\documentclass[pre,amsmath,amssymb,twocolumn,nofootinbib,floatfix]{revtex4-2}
\usepackage[colorlinks=true,linkcolor=blue,urlcolor=blue,citecolor=blue]{hyperref}
\usepackage{graphicx}
\usepackage{times}
\usepackage{color}

\usepackage{tikz}
\usetikzlibrary{arrows,shapes,positioning,decorations.pathmorphing}
\usetikzlibrary{decorations.markings}
\usetikzlibrary{snakes}
\tikzset{snake it/.style={decorate, decoration=snake,segment length=3mm}}
\tikzstyle arrowstyle=[scale=1]
\tikzstyle directed=[postaction={decorate,decoration={markings,
    mark=at position.65 with {\arrow[arrowstyle]{stealth}}}}]
\tikzstyle endreversedirected=[postaction={decorate,decoration={markings,
    mark=at position 1.0 with {\arrow[arrowstyle]{stealth}}}}]
\tikzstyle enddirected=[postaction={decorate,decoration={markings,
    mark=at position 1.0 with {\arrow[arrowstyle]{stealth}}}}]
\tikzstyle reverse directed=[postaction={decorate,decoration={markings,
    mark=at position.65 with {\arrowreversed[arrowstyle]{stealth};}}}]
\usetikzlibrary{decorations.markings}
\tikzset{->-/.style={decoration={
  markings,
  mark=at position #1 with {\arrow{>}}},postaction={decorate}}}

\definecolor{orange}{rgb}{1,0.6, 0}
\definecolor{darkergreen}{rgb}{0,0.7,0}
\definecolor{grey}{rgb}{.4,.4,0.4}

\newlength{\bilderlength}

\usepackage[normalem]{ulem}

\renewcommand{\doi}[2]{\href{http://dx.doi.org/#1}{#2}}
\newcommand{\arxiv}[1]{\href{http://arxiv.org/abs/#1}{#1}}
\newcommand{\link}[2]{\href{http://#1}{#2}}

\newcommand{\Eq}[1]{Eq.~(\ref{#1})}
\newcommand{\Eqs}[1]{Eqs.~(\ref{#1})}
\newcommand{\eq}[1]{(\ref{#1})}

\newcommand{\bea}{\begin{eqnarray}}
\newcommand{\eea}{\end{eqnarray}}

\newcommand{\black}{\color{black}}

\newcommand{\rme}{\mathrm{e}}
\newcommand{\rmd}{\mathrm{d}}
\newcommand{\nn}{\nonumber}
\renewcommand{\epsilon}{\varepsilon}
\newcommand{\ca}[1]{{\cal #1}}
\newcommand{\be}{\begin{equation}}
\newcommand{\ee}{\end{equation}}
\newcommand{\Fig}[1]{\includegraphics[width=\columnwidth]{./#1}} 

\graphicspath{{figures/},{}}

\newcommand{\us}{\cite{MukerjeeWiese2022}}

\begin{document}

\bibliographystyle{KAY-hyper}

\title{Depinning   in the quenched Kardar-Parisi-Zhang class I: Mappings, simulations and algorithm}
\author{Gauthier Mukerjee$^1$,  Juan A. Bonachela$^2$, Miguel A. Mu\~{n}oz$^3$, Kay J\"org Wiese$^1$}
  \affiliation{\mbox{$^1$ Laboratoire de Physique de l'\'Ecole Normale Sup\'erieure, ENS, Universit\'e PSL, CNRS, Sorbonne Universit\'e,} \mbox{Universit\'e Paris-Diderot, Sorbonne Paris Cit\'e, 24 rue Lhomond, 75005 Paris, France}\\
  \mbox{$^2$  Department of Ecology, Evolution, and Natural Resources, Rutgers University, New Brunswick, NJ, United States}
  \mbox{$^3$Departamento de Electromagnetismo y Física de la Materia and Instituto Carlos I de Física Teórica y Computacional,}\\
  \mbox{Universidad de Granada, Granada, Spain}}

\begin{abstract}

Depinning of elastic systems advancing on disordered media can usually be described by the quenched Edwards-Wilkinson equation (qEW). However, additional ingredients such as anharmonicity and forces that can not be derived from a potential energy may generate a different scaling behavior at depinning. The most experimentally relevant is the Kardar-Parisi-Zhang (KPZ) term, proportional to the square of the slope at each site, which drives the critical behavior into the so-called quenched KPZ (qKPZ) universality class.
We study this universality class both numerically and analytically: by using exact mappings we show that at least for $d=1,2$ this class encompasses not only the qKPZ equation itself, but also anharmonic depinning and a well-known class of cellular automata introduced by Tang and Leschhorn. We develop scaling arguments for all critical exponents, including   size and duration of avalanches. The scale is set by the confining potential strength $m^2$. This allows us to 
estimate numerically     these exponents as well as the $m$-dependent effective force  correlator $\Delta(w)$, and  
its correlation length   $\rho:=\Delta(0)/|\Delta’(0)|$.
Finally we present a new algorithm to  numerically estimate    the effective ($m$-dependent) elasticity $c$, and the effective KPZ non-linearity $\lambda$. This allows us to define a dimensionless universal KPZ amplitude ${\cal A}:=\rho \lambda /c$, which  takes the value ${\cal A}=1.10(2)$ in all systems considered in $d=1$. This proves that qKPZ is the effective field theory for all these models. Our work paves the way for a deeper understanding of depinning in the qKPZ class, and in particular, for the construction of a field theory that we describe in a companion paper.

\end{abstract} 

\maketitle

\section{Introduction}
Diverse systems can be modeled as an elastic object (line, surface, manifold) advancing through a random medium with quenched disorder:  disordered magnets and the associated Barkhaussen noise \cite{GiamarchiLeDoussal1995}, expanding fronts of bacterial colonies  \cite{BonachelaNadellXavierLevin2011,HuergoMuzzioPasqualeGonzalezBolzanArvia2014}, systems that show self-organized criticality \cite{AlavaMunoz2002}, or coffee imbibing this paper (if you are old fashioned enough to use a printout and to pour your coffee onto it) \cite{AmaralBarabasiBuldyrevHarringtonHavlinSadr-LahijanyStanley1995,PlanetDiaz-PiolaOrtin2020,HoltzmanDentzPlanetOrtin2020}.
Often, the elastic system experiences a so-called ``depinning transition" as a function of some driving parameter, so that the system changes from being pinned in some configuration to advancing at an average velocity \cite{Wiese2021}. 
 This phase transition can be    thought of as  
    the transition between an active state, where the elastic interface (or surface, etc)  changes over time, and an absorbing or quiescent state, where the interface remains frozen \cite{Hinrichsen2000}. At the transition, the dynamics becomes universal at sufficiently large scales, universality appears as microscopic details are irrelevant, and  different systems and models can be grouped together into a few  universality classes.
The latter can  then be studied via the renormalization-group (RG) and, more specifically, by employing functional renormalization group (FRG) approaches \cite{DSFisher1986,NarayanDSFisher1993a,NattermannStepanowTangLeschhorn1992,LeschhornNattermannStepanowTang1997,ChauveLeDoussalWiese2000a,LeDoussalWieseChauve2002,LeDoussalWieseChauve2003}.

 The simplest prototypical model for depinning transitions is the quenched Edwards Wilkinson equation (qEW), also called {\em harmonic depinning}. It monitors the height $u(x,t)\in \mathbb{R}$  of a $d$-dimensional interface embedded into $d+1$ dimensions.
 By construction, this excludes  overhangs as well as  bubbles. Its dynamics is described by  
\be
  \eta      \partial_t u(x,t) =\!\!  \underbrace{c  \nabla^{2}u(x,t)}_\text{{harmonic elasticity}}\!\!  {+} \underbrace{m^{2}[w{-} u(x,t)]}_\text{{confinement and driving}}  
        +  \underbrace{F\big(x, u(x,t)\big)}_\text{{pinning forces}}. \label{qEW}
\ee
The  pinning forces $F(x,u)$ are  quenched Gaussian random variables with 
variance $\overline{F(x,u)F(x',u')}= \delta^d(x-x') \Delta_0(u-u')$. $\Delta_0(u)$ is the  microscopic disorder-force correlator, assumed to decay rapidly for short-range (SR) disorder \cite{Wiese2021}.
The system is driven by slowly increasing $w$, either as $w=vt$ (with $v$ small), or via a small ``kick",  $w\to w+\delta w$ whenever the interface is stuck (pinned). 
The latter protocol is particularly useful to study avalanches \cite{SethnaDahmenMyers2001,RossoLeDoussalWiese2009a,LeDoussalWiese2008c,LeDoussalMiddletonWiese2008,LeDoussalMuellerWiese2010,LeDoussalWiese2009a,LeDoussalWiese2011b,LeDoussalMuellerWiese2011,LeDoussalWiese2012a,DobrinevskiLeDoussalWiese2013,DobrinevskiLeDoussalWiese2014a,ThieryLeDoussalWiese2015,AragonKoltonDoussalWieseJagla2016,DelormeLeDoussalWiese2016,DurinBohnCorreaSommerDoussalWiese2016,ThieryLeDoussalWiese2016,ZhuWiese2017,BonachelaAlavaMunoz2008}.

While the microscopic force correlator $\Delta_0(u)$ can be analytic,  the effective renormalized correlator $\Delta(w)$ seen in the field theory 
\cite{DSFisher1986,NarayanDSFisher1993a,NattermannStepanowTangLeschhorn1992,LeschhornNattermannStepanowTang1997,ChauveLeDoussalWiese2000a,LeDoussalWieseChauve2002,LeDoussalWieseChauve2003}, and measurable in experiments \cite{LeDoussalWiese2006a,MiddletonLeDoussalWiese2006,LeDoussalWieseMoulinetRolley2009,WieseBercyMelkonyanBizebard2019,terBurgBohnDurinSommerWiese2021} exhibits a cusp at $w=0$. The  slope at the cusp  is proportional to the typical avalanche size, $|\Delta'(0^+)|\sim \left<S^2\right>/\left< S\right>$ 
\cite{LeDoussalWiese2008c}.

The qEW class is not the only universality class for  interface depinning.
As we   show here,  there is one other rather large universality class, which we will establish is relevant whenever {\em non-linear effects cannot be neglected}.
As an example, the coffee imbibing our paper can be modeled by the cellular automaton
 proposed in 1992 by Tang and Leschhorn  (TL92) \cite{TangLeschhorn1992}, or variants thereof \cite{BuldyrevBarabasiCasertaHavlinStanleyVicsek1992,BuldyrevHavlinStanley1993}. 
As it permeates through the paper, the coffee  is blocked by a percolating line orthogonal to the coffee front, a phenomenon   known as  {\em directed-percolation depinning} (DPD) \cite{AmaralBarabasiBuldyrevHarringtonHavlinSadr-LahijanyStanley1995}. 
At a coarse-grained level one observes that the coffee front tends to grow in its normal direction, inducing anisotropy in the invaded medium. This phenomenon is modeled by an additional term, usually called {\em KPZ term}, \cite{KPZ} in the equation of motion, 
\be\label{KPZ-term}
 \eta \partial_t u(x,t) =... + \underbrace{ {\lambda}  [\nabla u(x,t)]^2}_\text{{KPZ-term}}.
\ee
In addition to fluid invasion (our coffee front) \cite{AmaralBarabasiBuldyrevHarringtonHavlinSadr-LahijanyStanley1995,PlanetDiaz-PiolaOrtin2020,HoltzmanDentzPlanetOrtin2020}, 
experiments on bacterial colonies \cite{BonachelaNadellXavierLevin2011,HuergoMuzzioPasqualeGonzalezBolzanArvia2014} or chemical reaction fronts \cite{AtisDubeySalinTalonLeDoussalWiese2014,ChevalierDubeyAtisRossoSalinTalon2017}  
share this property.  

In our setup, there is a preferred direction in the medium: the coffee front starts from a flat initial condition. If there is no such preferred direction, or the microscopic disorder is very strong,   the  critical behavior may be different \cite{Grassberger2020}.

Finally, the elastic restoring force  may be stronger than the harmonic elasticity  in \Eq{qEW}. This is particularly important at depinning in dimension $d=1$, where the roughness exponent $\zeta=5/4$ is larger than 1, meaning that the width of the interface grows stronger than the system size. 
As  argued in Ref.~\cite{RossoHartmannKrauth2002} this implies  that the small-displacement expansion for the elastic energy must be invalid, and one needs to go to the next order and include anharmonic elastic terms to  bring  the roughness to $\zeta\le 1$. 
The interpretation in Ref.~\cite{RossoHartmannKrauth2002} is that an elastic string would break and the qEW model is unphysical. For domain walls in 2d magnets this leaves two possibilities:  either they are self-affine interfaces in the qKPZ class, or non-self-affine (i.e.~not  described by a height function), and then possibly (isotropic)  fractals.

We   show below that all these models   belong to the same universality class,   termed the 
 {\em quenched Kardar-Parisi-Zhang} (qKPZ) universality class \cite{TangKardarDhar1995}. 
The field theoretic treatment of qKPZ via  FRG is, however, fraught with difficulties \cite{LeDoussalWiese2002}. 
The  reason is that in \cite{LeDoussalWiese2002} the effective KPZ coupling $\lambda$ generically flows to  strong coupling, indicating that the perturbative treatment breaks down. 
Ref.~\cite{LeDoussalWiese2002} further contained a subspace of fixed points defined by closed RG equations. This subspace is characterized by an exponentially decaying effective force  correlators $\Delta(w)$.  
 Our study was motivated by the observation that such a fixed point is indeed realized for the pair-contact process (PCP) \cite{Jensen1993}.
However, our numerical simulations indicate that none of the models discussed above has an exponentially decaying $\Delta(w)$.
 A new field theory that agrees with the simulations is therefore needed.
 
In view of the theoretical problems, here we tackle the system first numerically, and use the results to guide development of the theory. 
The first key conceptual advance is the introduction of a confining potential   proportional to $m^2$.
When $m^2=\infty$, the interface position is the flat configuration $u(x,t)=w$. As a consequence, both the elastic term $c \nabla^2 u(x,t)$ as the KPZ term $\lambda[\nabla u(x,t)]^2$ vanish. Thus when we sample  the total forces acting on the interface,  and its correlations $\Delta(w)$ (see section \ref{s:Effective action}), we see the microscopic correlations $\Delta_0(w)$ of $F(x,u)$. Let us now decrease $m^2$. As the interface   explores more  configurations and takes advantage of stronger pinning  configurations, the total pinning force increases,  while at the same time the interface becomes wider. 
Viewing the dynamics  as  a function of  $w$,   the size of jumps increases (with decreasing $m$), while their rate decreases. This leads to  larger   correlation lengths  $ \xi_m$ in the $x$-direction and $\xi_\perp$ in the driving direction.

While we can take $m$ smaller and smaller, we cannot take $m=0$ to start with,
as we can not even define a steady state. 
 However, when $\xi_m$ surpasses the system size, its effect on the (spatial) correlations of $u(x)$  becomes invisible. Thus for all practical purposes, we have reached the scaling limit.

Apart from the effective (total) force correlator $\Delta(w)$, we  numerically estimate   the flow of the parameters $c$ and $\lambda$ as a function of the confining potential strength $m^2$, to assess whether the effective non-linearity $\lambda$ 
flows to infinity as  it did in the field theory of  \cite{LeDoussalWiese2002}, or reaches a fixed point. 
In the latter scenario, we could hope to be able to {\em repair} the field theory.
Our overall goal is to identify the  {\em effective} field theory, i.e. the effective  large-scale theory, without   having to resort to  field theory techniques, as e.g.\ a diagrammatic expansion.

Our second key advance is to construct an algorithm   to   estimate  all  parameters of the effective field theory, as a function of $m$. 
Our  conclusion is that  an effective field theory with finite, $m$-dependent parameters   exists, and   it  has the form of the qKPZ equation.
More specifically, we   define a dimensionless effective KPZ amplitude  $\ca A$, 
\be \label{Adef}
  \mathcal{A}:= \lim_{m\to 0} \frac{\lambda}{c} \rho,
\ee
where $\lambda$ is the KPZ non-linearity in \Eq{KPZ-term}, $c$ the effective elasticity in \Eq{qEW}, and $\rho$ the correlation length of the   effective force correlator (for a precise definition see section \ref{The universal KPZ amplitude A}).  
Since the limit in \Eq{Adef} exists,  the theory remains valid  for the more common setting of the qKPZ equation  without an $m^2$-term. 
$\ca A$ should be viewed as the effective KPZ-nonlinearity in dimensionless units: It vanishes in qEW, and  we show numerically that $\ca A$ is the same for the TL92 cellular automaton, qKPZ and anharmonic depinning. This  supports  our claim that   (at least in $d=1$, $2$)  there is only one universality 
class with   $\ca A\neq 0$ which differs from qEW   with   $\ca A= 0$. The qKPZ fixed point is relevant even if the deviations from qEW in the microscopic model are small.

This paper is organized as follows: in Section \ref{sec:models} we describe  the models we use. We  then show   through mappings that   these  models are in the same universality class (section \ref{sec:mappings}). Section \ref{sec:scaling} is devoted to a scaling analysis, with the confining potential $m^2$ defining a new class of exponents.
How to   estimate numerically  the effective field theory is described in section \ref{s:Effective action}, first for  the force correlator (sections \ref{sec:scalingdelta}-\ref{s:Measuring Delta}), and then  the coupling constants (sections \ref{sec:algorithm}-\ref{The universal KPZ amplitude A}).
Brief conclusions are offered in section \ref{s:Conclusion}.
In a companion paper \us\, we show how to      obtain  the effective   field theory from a diagrammatic approach.

\section{Models}\label{sec:models}
In this section, we define three models. The  first two are described by a continuous equation, while the third one is a  cellular automaton model, i.e a discrete microscopic model.  We show in section \ref{sec:mappings} that they can all be mapped onto each other. 
 
 \subsection{QKPZ  equation}\label{s:qKPZ}
The quenched KPZ equation (qKPZ) is defined as
\bea 
\eta \partial_t u(x,t) &=& c\nabla^2 u(x,t) +  {\lambda} \left[ \nabla u(x,t)\right]^2  \nn\\
&&+ m^2\big[w-u(x,t)\big] + F\big(x,u(x,t)\big).\qquad 
\label{eq:qkpz}
\eea
Rescaling $u$, and $F(x,u)$, we could set $\lambda\to 1$. 
We prefer to not rescale the disorder, and thus $\lambda$ will change under RG. Invariant under these transformations is the sign of   $\lambda v$, i.e.\ $\lambda$ 
times the driving velocity $v=\rmd w/\rmd t$.
A positive sign defines what is called the positive qKPZ equation. The negative qKPZ equation exhibits a very different phenomenology, with the propagating front  spontaneously generating facets \cite{Sneppen1992,MogliaAlbanoVillegasMunoz2014,AtisDubeySalinTalonLeDoussalWiese2014}. 

Discretization of the KPZ term (second term on the r.h.s.\ of \Eq{eq:qkpz}) is not trivial, and the choice made for it  is  important.  
We use the discretization of Ref.~\cite{LeeKim2005}, 
\bea
&& u(x, t + \delta t)  - u(x, t) \nn\\
&& =  \delta t \Big\{m^2\Big[w-u(x, t) \Big] + F\big(x,   u(x, t) \big)   \nn  \\
 &&\qquad~~  + c \Big[ u(x+1, t)  + u(x-1, t)  -2 u(x, t) \Big]  + \nn \\
&&\qquad~~ +   {\lambda} \left[  \frac{u(x+1, t)   - u(x-1, t) }{2}  \right]^2  
\Big\}.
\label{eq:discreteqkpz}\\
&& \text{with } \lambda = 3, \quad c=1, \quad \delta t = 0.01.
\label{eq:discreteqkpzparams}
\eea
Our main control parameter  $m$ is varied between  $0.05$ and $0.6$. The system size is chosen to be  $L \leq 512$ with $L$ the linear size. Following standard approaches, the disorder forces are drawn  from a Gaussian distribution with unit variance,   linearly interpolated between   integer values of $u$.
While efficient algorithms exists for the  other two models, a direct simulation of the qKPZ equation is  computationally expensive, and   we have restricted our simulations to $d=1$.

\subsection{Anharmonic depinning}
Anharmonic depinning (aDep) is defined by the     equation 
\bea
   \eta     \partial_t u (x,t)&=& \underbrace{c_4 \nabla \cdot  \left\{ \nabla   u(x,t) \big[ \nabla u(x,t)\big]^2 \right\}}_\text{{anharmonic elasticity ~}} + \underbrace{c  \nabla^{2}u(x,t)}_\text{{~harmonic elasticity}} \nn  \\
        &&    + \underbrace{m^{2}[w{-} u(x,t)]}_\text{{~driving force}}  +   \underbrace{F\big(x, u(x,t)\big)}_\text{{~quenched disorder}}. 
        \qquad 
        \label{eq:anhdep}  
\eea
If $\vec \rme_i $ represents the unit vectors in   $d$ dimensions, the   discretized anharmonic energies are
\bea
\label{Hel}
\ca H_{\rm el}[u] &=&  \sum_x\sum_{i=1}^d \ca E_{\rm el}\big ( u(x{+}\vec \rme_i) - u(x)\big), \\
\ca E_{\rm el}(u) &=& \frac c 2 u^2 + \frac {c_4}4 u^4.
\label{Eel}
\eea
 The resulting elastic forces at site $x$ are  
\bea  \label{eq:anhdepdiscrete} 
&&-\frac{\delta \ca H_{\rm el}[u]}{\delta u(x)} = \sum_{i} \ca E_{\rm el}'\big ( u(x{+}\vec \rme_i) {-} u(x)\big) 
+ \ca E_{\rm el}'\big ( u(x{-}\vec \rme_i) {-} u(x)\big) \nn\\
&&= 
 c \sum_{i=1}^d \Big[ u(x+\vec \rme_i)+u(x-\vec \rme_i)-2 u(x)\Big] \\
&&+ \,  {c_4} \sum_{i=1}^d   \Big[ u(x{+}\vec \rme_i) {-} u(x)\Big]^3 {+}  \Big[u(x{-}\vec \rme_i) - u(x)  \Big]^3.
 \nn
\eea\begin{figure}
\centering
\setlength{\unitlength}{1cm}
{{\black
\begin{picture}(8.7,6)
\put(0,0){\includegraphics[width=8.6cm]{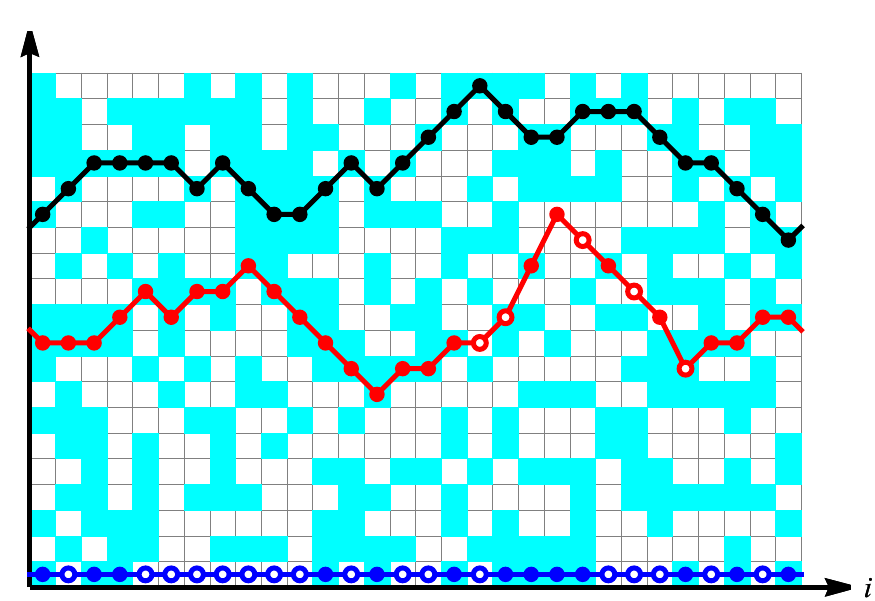}}
\put(8.2,0.35){$x,i$}
\put(0.3,5.6){$u$}
\end{picture}}}
\caption{The cellular automaton    TL92.  Blocking cells, i.e.\ cells above the threshold are drawn in  cyan; those  below in white. The initial configuration is the string at height 1 (dark blue). The interface moves up.  An intermediate configuration is shown in red,   the final configuration  in black. Open circles represent unstable points, i.e.\ points which can move forward; closed circles are stable.}
\label{fig:TL92_time_evol}
\end{figure}The  discretizations are similar to the  qKPZ equation.
For $d=1 $, we simulate systems with size up to $L = 2048$, assuming $ m \leq 0.05$. Using GPU in $d=2$ allowed us to reach $L=256$, with $m=0.09$. In  $d=3$, we reached   $L = 64$, and  $m= 0.08$.  
To speed up simulations, we set   
$c = 0$, after checking that it gives the same results as $c  = 1$.
We varied the anharmonic term as
   $c_{4} = 0.1, 0.2, 0.3$. There are  $6\times 10^{5}$ ($d=1$), $5\times 10^{4}$ ($d=2$), and  $2\times 10^{4}$ ($d=3$) independent samples. 
By construction, this system only moves forward,  respecting the Middleton ``no-passing" theorem \cite{Middleton1992}, see section 
\ref{s:No-passing theorems}. This allows us to use the very efficient algorithm  introduced in Ref.~\cite{RossoKrauth2001b}.

\subsection{TL92}
\label{s:TL92}

To describe fluid imbibition, Tang and Leschorn introduced 
the  cellular automaton visualized in Fig.~\ref{fig:TL92_time_evol}  (TL92)  \cite{TangLeschhorn1992}. 
On  a square lattice, each cell $ (i,j)$ is assigned a variable $f(i,j) \in  [0,1]$.  If $ f(i,j) < p$  the cell is considered closed (blocking). Otherwise the cell is considered open. 
 The interface starts from a flat configuration at the bottom (dark blue on Fig.~\ref{fig:TL92_time_evol}). A point $(i, j=u_i)$ on this interface is unstable and can move forward by 1, $u_i \rightarrow u_i + 1,$  according to the following rules (in that order): 
\begin{itemize} 
\item[(i)] links cannot be longer than 2. If a site is 2 cells ahead of its neighbors, stop. 
\item[(ii)] if $ f(i,j) > p$,  move forward. 
\item[(iii)] if one of the neighboring sites is 2 cells ahead, move forward.
\end{itemize}
While in the original version the critical force $p$ is a constant  \cite{TangLeschhorn1992}, 
Here we choose it to depend on the height $j=u_i$ of the interface,
\be\label{fc-i}
 p:= m^2 [u_i-w].
\ee
This has two consequences: first, as $f(i,j)\in [0,1]$,  rule (ii) is satisfied for all $u_i<w$, and never satisfied for $u_i>w+m^{-2}$.
As a result, the   interface is trapped in a domain of size $m^{-2}$. Increasing $w$, we can drive the interface as in qEW, \Eq{qEW}.
Our protocol is to keep $w$ fixed until a stable configuration is reached, and only then increase  $w$ by $\delta w $. Two   timescales are thus separated: a fast one governing the reorganization of the system, and an infinitely slower one corresponding to the driving. We use this   protocol   to calculate the effective force  correlator and to simulate  avalanches.

The interface is pinned when all its cells are blocking; its maximal slope is 1, see Fig.~\ref{fig:TL92_time_evol}. 
This ensures that a directed-percolation path goes from left to right, i.e.\ perpendicular to the driving direction  \cite{TangLeschhorn1992}. To be precise, the line gets pinned at the lowest percolating cluster (see Fig.\ \ref{f:dir-percol2} below). 
This relation to directed percolation allows us to use many of the high-precision results available for DP (see section \ref{s:Connection to directed percolation}). 
Since time in the DP formulation is from left to right, whereas time for depinning of an interface is in the $u$ direction, 
this correspondence is restricted to static quantities i.e. those corresponding to blocking configurations. 

In our simulations of this cellular automaton, we use $L  = 4096$ and $m\geq0.02$  in $d=1$. GPU computing allows us  to reach $L=256$ at $m=0.05$ in $2d$, and $L=128$ and $m=0.1$ in $3d$.

\section{Mappings}\label{sec:mappings}
In this section we present different mappings between the three models introduced in sections \ref{s:qKPZ} to \ref{s:TL92}. Some of these mappings rely on no-passing theorems, which we prove below first.

\subsection{No-passing theorems for TL92 and anharmonic depinning}
\label{s:No-passing theorems}

In parallel to the famous no-passing theorem by  Middleton \cite{Middleton1992} for harmonic depinning, we now prove a similar theorem for TL92 and anharmonic depinning.

\subsubsection{No-passing theorem for anharmonic depinning in the continuoum}

\noindent{{\bf Assumptions.}

\noindent $\bullet$  the elastic energy  between nearest neighbors $\ca E(u)$ in \Eq{Eel} is a convex function, 

\noindent $\bullet$ the disorder force $F(x,u)$ is continuous in $u$,

\noindent $\bullet$  $\dot u(x,t)\ge 0$.

\smallskip

\noindent{{\bf No-Passing Theorem I (depinning in the continuuum).}}

\noindent $\bullet$ $\dot u(x,t')\ge 0$ for all $t'\ge t$. 

\noindent $\bullet$   if two configurations are ordered, $u_2(x,t)\ge u_1(x,t)$, then they remain ordered for all times, i.e.\ $u_2(x,t')\ge u_1(x,t')$ for all $ t'>t$.

\smallskip
\noindent{{\bf   Proof.}} Consider an interface discretized in $x$. 
The trajectories $
u(x,t)
$ are a function of time.  Suppose that there exists $x$ and $t'>t$  such that $\dot u(x,t')<0$. Define $t_0$ as the first time when this happens, 
$t_0:=\inf_x\inf_{t'>t}\{ \dot u(x,t')<0\}$, and $x_0$ the corresponding position $x$. By continuity of $F$ in $u$, the velocity $\dot u$ is   continuous in time, and  $\dot u(x_0,t_0)=0$. 
This implies that the disorder force acting on $x_0$  does not change in the next (infinitesimal) time step, and the only changes in force can come from a change in the elastic terms. Since by assumption no other point has a negative velocity and the elastic energy   $\ca E(u)$ is convex, this change in force can not be negative,  contradicting   the assumption. 

To prove the second part of the theorem, 
consider the following configuration at time $t_0$:
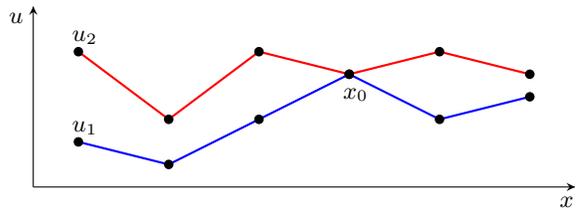
\begin{figure}[t]
{\parbox{7.5cm}{{\begin{tikzpicture}[scale=1.2]
\coordinate (v1) at  (0,0) ; 
\coordinate (v2) at  (1,-.25) ; 
\coordinate (v3) at  (2,.25) ; 
\coordinate (v4) at  (3,.75) ; 
\coordinate (v5) at  (4,.25) ; 
\coordinate (v6) at  (5,.5) ;  
\coordinate (u1) at  (0,1) ; 
\coordinate (u2) at  (1,.25) ; 
\coordinate (u3) at  (2,1.) ; 
\coordinate (u4) at  (3,.75) ; 
\coordinate (u5) at  (4,1) ; 
\coordinate (u6) at  (5,.75) ;  
\node (x) at  (5.4,-.5)    {$\!\!\!\parbox{0mm}{$\raisebox{-3mm}[0mm][0mm]{$ x$}$}$};
\node (x0) at  (3,0.7)    {$\!\!\!\parbox{0mm}{$\raisebox{-3mm}[0mm][0mm]{$ x_{\rm 0}$}$}$};
\node (uu1) at  (0,0)    {$\!\!\!\parbox{0mm}{$\raisebox{1mm}[0mm][0mm]{$ u_1$}$}$};
\node (uu2) at  (0,1)    {$\!\!\!\parbox{0mm}{$\raisebox{1mm}[0mm][0mm]{$ u_2$}$}$};
\node (u) at  (-.7,1.2)    {$\!\!\!\parbox{0mm}{$\raisebox{1mm}[0mm][0mm]{$ u$}$}$};
\draw [blue,thick] (v1) -- (v2)--(v3)--(v4)--(v5)--(v6);
\draw [red,thick] (u1) -- (u2)--(u3)--(u4)--(u5)--(u6);
\draw[enddirected](-.5,-.5)--(5.5,-.5);
\draw[enddirected](-.5,-.5)--(-.5,1.5);
\fill (v1) circle (1.5pt);
\fill (v2) circle (1.5pt);
\fill (v3) circle (1.5pt);
\fill (v4) circle (1.5pt);
\fill (v5) circle (1.5pt);
\fill (v6) circle (1.5pt);
\fill (u1) circle (1.5pt);
\fill (u2) circle (1.5pt);
\fill (u3) circle (1.5pt);
\fill (u4) circle (1.5pt);
\fill (u5) circle (1.5pt);
\fill (u6) circle (1.5pt);
\end{tikzpicture}}}}
\caption{Two configurations at depinning.}
\label{2confifs}
\end{figure}
Here the red configuration is ahead of the blue one, except at position $x_0$, where they coincide. 
As in the first part of the proof, we wish to bring to a contradiction the hypothesis that at some later time  $u_1(x_0)$ (blue in Fig.~\ref{2confifs}) is ahead of $u_2(x_0)$ (red). For this reason, we have chosen $t_0$ the infimum of   times contradicting the theorem, $t_0:=\ \inf_{t'>t}\{  u_1(x_0,t')>u_2(x_0,t')\}$.
 Consider  
the    equation of motion for the difference between $u_1$ and $u_2$ at point $x_0$. According to \Eqs{eq:anhdep} and  more generally \eq{eq:anhdepdiscrete}, 
\bea \nn
&\eta &\partial_t \left[ u_2(x_0,t) - u_1(x_0,t)\right]\big|_{t=t_0}\\
&& = \sum_{x= {\rm nn}(x_0,t_0)} \ca E_{\rm el}'\big ( u_2(x,t_0) -u_2(x_0,t_0)\big) \nn\\
&& \qquad\qquad \quad-\ca E_{\rm el}'\big ( u_1(x,t_0) - u_1(x_0,t_0)\big)  \nn\\ 
&& \ge 0.
\eea
The sum runs over all nearest neighbours of $x_0$. 
The disorder forces have canceled as well as the term of order $m^2$, since by assumption $u_2(x_0,t_0) = u_1(x_0,t_0)$. The inequality follows from   monotonicity of  $\ca E'_{\rm el}(u)$, equivalent to the convexity of $\ca E(u)$.

\subsubsection{No-passing theorem for anharmonic depinning as a cellular automaton}

\noindent{{\bf Assumptions.}}

\noindent  $\bullet$ the positions $u(x,t)$ are integers; they can grow by $1$ in an update, or remain constant.

\noindent $\bullet$  the elastic energy  between nearest neighbors $\ca E(u)$ is convex.

\smallskip

\noindent{{\bf No-Passing Theorem II (cellular automaton for depinning)}:}

\noindent $\bullet$   if two configurations are ordered, $u_2(x,t)\ge u_1(x,t)$, then they remain ordered for all times, i.e.\ $u_2(x,t')\ge u_1(x,t')$ for all $ t'>t$.

\smallskip
\noindent{{\bf   Proof.} As for theorem I.

\subsubsection{No-passing theorem for TL92}

\noindent{{\bf No-Passing Theorem III (TL92)}.}

\noindent $\bullet$   if two configurations are ordered, $u_2(x,t)\ge u_1(x,t)$, then they remain ordered for all times, i.e.\ $u_2(x,t')\ge u_1(x,t')$ for all $ t'>t$.

\smallskip
\noindent{{\bf   Proof.} Consider again point $x_0$ on Fig.~\ref{2confifs}. By direct inspection of all possible configurations one shows that either nobody moves, red and blue move together, or only red moves. 
This works as well for parallel update as for single-site update, provided one updates both configurations at the same time.

\subsection{Mapping from anharmonic depinning to TL92}\label{sec:mapanhTL92}

\subsubsection{General idea}

The general idea is to show that, {\em for an appropriate choice of parameters}, both TL92 and anharmonic depinning have the same blocking configurations. This statement has two directions:

\noindent  $\bullet$ a blocking configuration for TL92 is   a blocking configuration for the  depinning of an elastic line. 

\noindent $\bullet$  all other configurations move forward. 

We failed to prove the stronger statement,  namely that each site stable in an allowed TL92 configuration  is also stable for depinning, and that only allowed TL92 configurations appear at depinning. This means that the dynamics of both models may be different, and even show different dynamical critical exponents.

\subsubsection{Cellular automaton in $d=1$}
\label{Cellular automaton in d=1}

We aim to find the blocking configurations of    TL92    with an interface whose law of evolution is the one of anharmonic depinning as given in \Eq{eq:anhdep}, at least with a specific choice of parameters. 
To that end,  we need to check for a given disorder that every configuration of the interface  following the anharmonic depinning equation stops   at the same configuration as in  TL92.  
Let us  start with a cellular automaton version.
 We choose disorder forces $F = \pm1$, where $F=1$ corresponds to open sites, and $F=-1$ to blocking sites.
 The height is integer, and whenever a site is unstable, it is moved ahead by $1$, as in TL92. 
 Whether a site moves or not depends only on its relative position to its nearest neighbors. Therefore, we only need to test whether there is agreement for 25  configurations: the two neighbors of a site can be  separated by a distance with values in  $\{-2, -1, 0, 1,2\}$. The maximum distance is given by the  TL92 rule that  two neighbors can not be separated by a distance of more than $2$, a condition anharmonic depinning needs to satisfy too.
Symmetry of the forces under the exchange of the left and right neighbors  decreases the number of distinct cases to  15. 
  If one can find parameters  $c$ and $c_4$,  such that the two prescriptions agree on those 15 configurations, the anharmonic depinning equation    agrees for any interface configuration with TL92, and we   have our mapping.

For each configuration to be tested, there are three sites to check: the left, the middle  and the right ones. We 
note the relative position of the left and right neighbors: for example $ (+1,+1) $ corresponds to a ``v" shape with slope $1$, $ (-2,+2) $ to ``/'' with slope $2$, and so on. 
Our considerations are for the middle point. 
If  according to the  TL92  rules it moves, the force felt by it must be $\geq 0$, otherwise $\leq 0$.
The   discretization of the equation of motion is given in  \Eq{eq:anhdepdiscrete}, with elastic forces between two neighbors  $c_4 (u_i -u_{i+1})^3$.  
This gives the inequalities  in  Table \ref{tab:mapAnh1d}. With one exception, they are fulfilled  by taking $c_4 \in ]\frac1 7, \frac1 2[$; a centred value of $c_4=1/4$ is a good choice. 
Then in all cases anharmonic depinning has the same update rules as TL92, except for the configuration  $(2,-2)$, a strongly inclined line. We did not succeed to tweak the model such that this configuration is also always stable at depinning. On the other hand, all blocking TL92 configurations are also blocking at depinning, and there is no configuration blocked at depinning which would move in TL92.
\begin{table}[t]
\centering
\begin{tabular}{|cc|cc|}
\hline
\textrm{configuration}&
\textrm{condition} &
\textrm{configuration}&
\textrm{condition} \\
\hline
$(2,2)$ & $  16 c_4>1 $ & $(1,-2)$&  $7 c_4 >1$ \\
$(2,1)$ &$ 9 c_4   > 1 $ & $(0,0)$ & true \\
  $(2,0)$&  $ 8c_4  > 1 $ &$(0,-1)$& $-1< c_4<1 $\\\
 $(2,-1)$& $ 7 c_4 >1  $ &$(0,-2)$ &$ 8 c_4 >1$ \\
 $(2,-2)$& false &        $(-1, -1)$&$ -1< 2c_4<1 $\\
 $(1,1)$& $-1< 2c_4 <1   $& $  (-1,-2)$&$ 9c_4>1 $\\
$(1,0)$& $-1<  c_4 <1  $&$(-2,-2)$&$ 16 c_4>1 $\\
 $(1,-1)$& true && \\
 \hline
\end{tabular}
 \caption{Conditions on the anharmonic depinning coefficients,  such that anharmonic depinning evolves as TL92, for each of the configurations in TL92.}
 \label{tab:mapAnh1d}
\end{table}

Using the no-passing theorems II and III shows that both models have exactly the same blocking configurations, and that they are chosen independently from the history. 
Since anharmonic depinning can move faster than TL92, we conclude that their corresponding dynamical exponents should satisfy
\be
z_{\rm TL92} \ge z_{\rm aDep}.
\ee

\subsubsection{Cellular automaton in an arbitrary dimension}
\label{Cellular automata in arbitrary dimension}
We now generalize our considerations to an arbitrary dimension $d$. We first derive necessary conditions for a (globally) blocking configuration in TL92.
\begin{itemize}
\item[(i)] A blocking configuration of TL92 has no site whose neighbor is at a distance $-2$. 
\item[(ii)] A blocking configuration of TL92 has no site whose neighbor is at a distance $2$. 
\end{itemize}
As (ii) is trivial, we only need to 
prove (i): Suppose   a site $s_1$ exists with a neighbor $s_2$ at a distance $-2$. Then their heights  $u(s)$ satisfy 
\be
u(s_2) = u(s_1)-2.
\ee
If site $s_2$ has a neighbor $s_3$ which is at a height distance $-2$, we continue to $s_3$, and so on. Since $u(s_i)$ is a decreasing sequence, and the minimum of all heights $u_{\rm min}:=\min_s u(s) $ exists, this process stops, say at step $n$.
By construction  site $s_n$ has no neighbor at distance $-2$, but at least  neighbor $s_{n-1}$ at distance $2$. Thus site $s_n$ is unstable,   proving (i).

Let us now check for each site $s$ its local configuration $l_s =(\delta u_1,...,\delta u_{2d})$, defined as in section \ref{Cellular automaton in d=1}. 
We start with a globally blocking configuration in TL92. Due to (i) and (ii), all its $\delta u_i \in \{-1,0,1\}$. Whether the site is stable or not is  {\em disorder decided}. We have to ensure that this is the same for anharmonic depinning. To simplify matters, we set $c\to 0$, only retaining $c_4$. 
In TL92 the site is unstable if $F=1$ and stable if $F=-1$.  
Let us consider the stable case. 
In order to reproduce this in anharmonic depinning, we need that even if all neighbors pull in the opposite direction, i.e.\ are $1$, the site remains stable. 
This implies the
\begin{itemize}
\item[(iii)] condition from configuration $l_s=(1,1,....,1)$:
 \be
c_4 < \frac 1{2d}.
\label{15}
\ee
\end{itemize}
If the site in TL92 is unstable, the same condition arises, this time for the configuration $(-1,-1,...,-1)$.

Next consider a configuration with one $-2$:
\begin{itemize}
\item[(iv)] condition from $l_s=(-2,...)$: none.
\end{itemize}
TL92 is blocked, while aDep may move or not. Due to the no-passing theorems, nothing has to be checked. 

Remains to check a configuration with at least one $2$: 
\begin{itemize}
\item[(iv)] condition from $l_s=(2,...)$:
\be\label{c4lower}
c_4 ({9-2d} ) > 1.
\ee
\end{itemize}
Proof: We need aDep to move. The worst case is that the disorder is $F=-1$, and that all remaining neighbors pull backwards. Since we have already excluded case (iv), they can maximally be at a distance $-1$. This gives the condition that the total elastic  force $c_4[ 2^3 -(2d-1)] >1$. Simplifying yields \Eq{c4lower}.

We conclude that TL92 and aDep always find  the same blocking configurations (in $d\le 4$), as long as  
\be
\frac{1}{9-2d} < c_4 < \frac 1{2d}.
\ee
This gives the bounds
\bea
\frac{1}{7} < c_4^{d=1} < \frac 1{2},\\
\frac{1}{5} < c_4^{d=2} < \frac 1{4}.
\eea
In $d=3$ there is no solution, but one can repeat the argument with an anharmonicity 
\be
\ca E_{\rm el}(u)= \frac{c_{2p}}{2p} u^{2p}.
\ee
While the bound \eq{15} remains valid as a condition for $c_{2p}$, \Eq{c4lower} changes to
\be
c_{2p}(2^{2p-1}+1-2d)>1.
\ee
Therefore the simplest solution in $d=3$ reads
\be
 \frac1{ 27}<c_6 < \frac 1 6.
\ee 
This leaves open the possibility that in $d=3$ several anharmonic-depinning universality classes exist. Both our simulations and the literature \cite{AmaralBarabasiBuldyrevHarringtonHavlinSadr-LahijanyStanley1995}  are in favor of that hypothesis.

}

\subsubsection{Depinning in the continuum}
In \cite{Wiese2021} (section 5.7) a continuum model was proposed in $d=1$ which finds the blocking configurations of TL92, and otherwise moves. The idea is to let the disorder act only close to integer values of $u$, and to provide a sufficiently strong force in between. This way anharmonic depinning stops at TL92 configurations, but never in between.

\subsection{Mapping qKPZ to TL92}\label{sec:mapqkpzTL92}
\begin{table}[t]
\centering
\begin{tabular}{|ccc|ccc|}
\hline
\multicolumn{3}{|c|}{TL92 blocking}&\multicolumn{3}{|c|}{TL92 non-blocking} \\
\hline
~configuration&
\textrm{condition} & &
\textrm{~configuration}&
\textrm{condition}&   \\
\hline
$(1,1)$ & $c <\frac12 $			& $\checkmark$			 &(2,2)& $c> \frac14 $ & $\checkmark$\\
$(1,0)$ &$   c +\frac \lambda 4<1 $ & $\checkmark$		& $(2,1)$ & $3 c + \frac\lambda 4 > 1$ & $\checkmark$\\
$  (1,-1)$&  $   {\lambda} < 1  $ 	& $\checkmark$		& $(2,0)$ & $2 c + \lambda > 1$  & $\checkmark$ \\
$ (0,0)$&  true  				& $\checkmark$			&$ (2,-1)$&$ 4 c+9 \lambda>4$  & $\checkmark$ \\
$ (0,-1)$&$  -1 < c{-} \frac{\lambda}4 <  1  $ & $\checkmark$ & $(2, -2)$ &$ 1+4 \lambda<0$& \\
$(-1,-1)$& $      c <\frac12  $      & $\checkmark$ & 			$(1,-2)$ 	&$ 4+9 \lambda<4 c $ & \\
& & &	$(0,-2)$  & $1+\lambda<2 c $ &   \\
& & & 	$(-1,-2)$ & $4+\lambda<12 c$ & $\checkmark$ \\
& & &	$(-2,-2)$ & $c > \frac1 4  $ & $\checkmark$\\
\hline
\end{tabular}
 \caption{Set of conditions on the qKPZ coefficients for each possible configuration in TL92. $F=1$ is  the maximum disorder force. We can satisfy most conditions by choosing $c=2/5$, $\lambda=1/2$, as indicated by the checkmarks. }
 \label{tab:mapqKPZ}
\end{table}

The mapping of qKPZ onto TL92 is more delicate as qKPZ has no  no-passing theorem. On Fig.~\ref{tab:mapqKPZ} we show the conditions to be satisfied for a cellular-automaton version of qKPZ, termed qKPZc. Inspection shows that not all conditions can  be satisfied simultaneously. This remained true if we enlarged the space of allowed models. As an example, we allowed for an additional constant term in the equation of motion.  

What we could however achieve,  is that blocking configurations of TL92 are blocking for qKPZc, while most of the non-blocking configurations of TL92 are non-blocking for qKPZc, choosing
\be\label{23}
c= \frac25, \quad \lambda = \frac12.
\ee
The violating cases $(2,-2)$, $(1,-2)$ and $(0,-2)$ do not move in TL92, but   move in qKPZc, bringing us out of the allowed configurations of TL92.

We now provide heuristic arguments   that  in the continuous version these configurations are not reached.
In the continuum and close to the depinning transition, we have $\frac{\xi_{\bot} }{\xi_m} \rightarrow 0$. As a result, at large distances   compared to the lattice cutoff but below the correlation length $\xi_m$, the interface must be flat on average. Now suppose that
a series of sites are aligned and has the maximal available slope $\alpha$. 
This extremal perturbation is shown in Fig.~\ref{fig:configqKPZTL92}.
We use the discretization of \Eq{eq:discreteqkpzparams}. We do not consider the disorder for simplicity. (It only enters in this argument through the structure of the space swept by the interface between avalanches.)
\begin{figure}
\includegraphics[width=\linewidth]{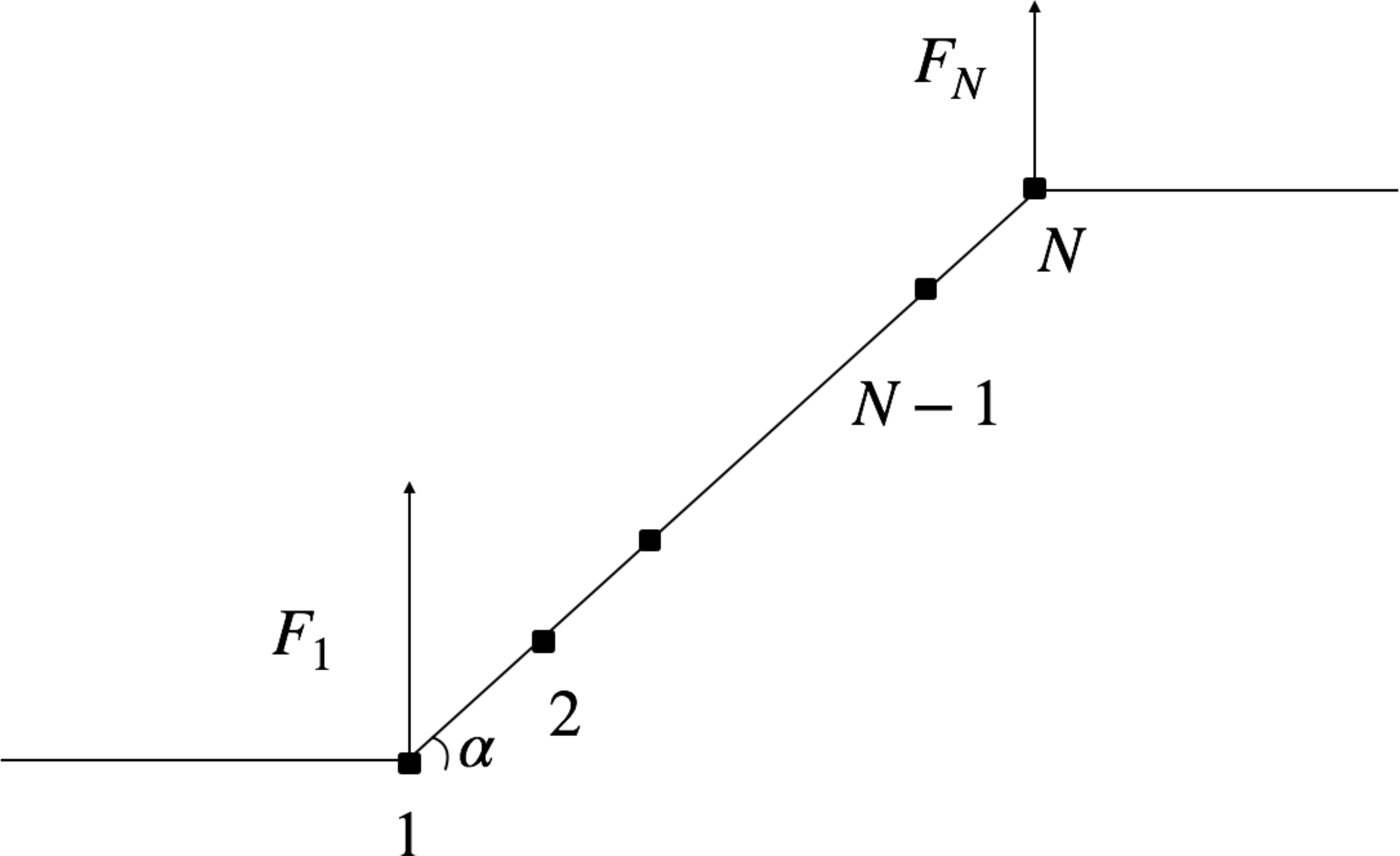}
\caption{ Possible instability of the qKPZ equation, when a serie of locally aligned points have a slope $\alpha$. The interface goes back to being flat at large distance, due to the $\xi_{\bot}  / \xi_m \rightarrow 0 $. The force felt by the lowest point is greater than the force felt by the lowest point. Below a certain slope $\alpha$, this configuration do not generate a local slope above $\alpha$.   }
\label{fig:configqKPZTL92}
\end{figure}
 If we name $F_1$ the force felt by the leftmost site of the slope, and $F_N$ the force felt by the rightmost of the slope, then (independent of the KPZ term)\be
\Delta F_{1N} = F_{N}- F_{1} = -2c{\alpha}.
\ee
As a result globally the perturbation gets flattened.
For $i \in \{2, 3,...,N-2 \} $ we have $\Delta F_{i, i+1} = 0 $. And finally
\bea
\Delta F_{12} = - \alpha c + \frac{3 \lambda \alpha^2}{4}, \\
\Delta F_{N-1, N} = - \alpha c -  \frac{3 \lambda \alpha^2}{4}. \label{eq:topofinstability}
\eea
The local slope does not increase if $\alpha \leq \frac{4 c}{3\lambda}$.
With the previous values for $\lambda$ and $ c$ this gives $\alpha \leq \frac{16}{15} $. 
So while we expect slope $\alpha=1$ to be commonly reached, larger slopes are not.   
As a result, the configurations $(2,-2)$ and $(1,-2)$ are not reached, and   that their associated conditions are not fulfilled   does not matter. Finally, for the case $(-2,0)$, it corresponds to the highest site of the perturbation. We can see from \Eq{eq:topofinstability} that it always experiences a force that is weaker than the site just below, and as a result the local slope gets flattened and can not reach the value $(-2,0)$.
Adding disorder is statistically more likely to pin the site $N$, which feels a weaker forward force, than the site $1$, accelerating the smoothening of the perturbation.
We checked  by numerical simulations of 
\Eqs{eq:discreteqkpz}-\eq{eq:discreteqkpzparams} in $d=1$,  with forces set to   $F=1$ for an open cell, and $F=-1$ for a blocking cell, and $c=\frac25$, $\lambda= \frac12 $ that qKPZ stops at the same  blocking configuration as TL92.

\subsection{Mapping anharmonic depinning to qKPZ}
Starting from the anharmonic-depinning \Eq{eq:anhdep}, the KPZ term is generated under renormalization, even in the limit of a vanishing driving velocity,  $v\rightarrow 0^+$, which corresponds to  depinning, under the combined action of the anharmonicity and the non analytic disorder force correlator. This was shown in  \cite{LeDoussalWiese2002}, and is reproduced in the companion paper \us.
The KPZ term generated is then more \emph{relevant},  in the renormalization-group sense, than the anharmonic elastic terms. This ensures that anharmonic depinning belongs to the qKPZ class.

\section{Critical exponents and scaling relations} \label{sec:scaling}
 
Some scaling relations have already been   derived \cite{TangLeschhorn1992,BuldyrevBarabasiCasertaHavlinStanleyVicsek1992,AmaralBarabasiBuldyrevHarringtonHavlinSadr-LahijanyStanley1995}, using 
the distance to the critical point as a control parameter. 
However, in order to construct the field theory, one has to introduce an  infrared regularization acting in the $x$-direction. This is achieved by driving the surface using  a confining potential with strength $m^2$, i.e.\ the term ${m^{2}[w- u(x,t)]}$ in \Eqs{qEW}, \eq{eq:qkpz}, \eq{eq:anhdep} and ${m^{2}[u_i-w]}$ in \Eq{fc-i}. It is this term which forbids  rare large fluctuations in the $u$-direction. Moreover, this term is crucial for the field theory to have a fixed point \cite{MukerjeeWiese2022},     to estimate  the effective-force correlations (see section \ref{s:Effective action} below), and to quantitatively compare the RG flow between field theory and simulations. It is thus necessary to derive all scaling relations in terms of the confining potential strength $m^2$, or {\em mass} $m$.  While the $m$-dependence  in correlation functions disappears for $ \xi_m \gtrsim  L$, having a finite (even tiny) $m$ allows us to be in the steady state.
All scaling relations are summarized in table \ref{t:scaling-relations}.

\begin{table}
\centering
\begin{tabular}{|c c c|}
\hline
&&\\
$  \xi_{\perp} \sim \xi_{\parallel}^\zeta$, &$ \quad \zeta= \frac{\nu_\perp}{\nu_\parallel}$, & $\zeta_m = \frac{2\nu_\bot}{1+ \nu_\bot},$\\
&& \\
$\frac{\zeta_m}{\zeta} = \nu_{\parallel}(2-\zeta_m)$,  &$\xi_m  = C m^{-\frac{\zeta_m}{\zeta}}$, &$\beta = \frac{\zeta_m(z-\zeta)}{\zeta(2-\zeta_m)}$, \\
&& \\
$\Psi_\eta  = z\frac{\zeta_m}{\zeta} - 2,$& $\Psi_{c} = 2 \frac{\zeta_m - \zeta}{\zeta}$, & $\Psi_\lambda = 2 \frac{\zeta_m - \zeta}{\zeta} - \zeta_m$,\\
&& \\
$\tau=2-\frac{2}{d \frac{\zeta_{m}}{\zeta}+\zeta_{m}}$,  & $(1 - \tau ){\frac{d+ \zeta}{z}}= 1- \alpha $. &\\
&&\\
\hline
\end{tabular}
\caption{All scaling relations derived in this paper. $\nu_\perp,\nu_\parallel $ come from DP mappings.}
\label{t:scaling-relations}
\end{table}

\subsection{Why use a confining potential of strength $m^2$?}\label{sec:m_protection}
The reader may wonder why we use a protocol with a confining potential of strength $m^2$, as in \Eqs{qEW}, and  \eq{eq:qkpz}-\eq{eq:anhdep} There are several reasons: 
First of all, this allows us to reach a steady state, and not only to approach it, as is the case when tuning an applied force to the depinning threshold. Second, 
having an energetically preferred position $w$ allows us to talk about fluctuations of the center of mass $u_w$ around $w$. 
Finally, as  discussed in section \ref{s:Effective action}, this allows us to measure correlations of the effective force. 

It is important that 
the confining-potential strength $m^2$ is protected, i.e.\   changes with scale $m$ as $m^2$ (without any correction):  On average the center of mass $u_w$ of the interface follows the driving term $w$, i.e.\ $m^2 \overline{[u_w-w]}=f_c$. If one changes
  $w\rightarrow w + \delta w$, this results in an increase in force $\delta f= m^2 \delta w$, and the center of mass  in the long-time limit is $u_w \to u_w+  \delta w$,   on average. As a result, the long-time response of the center of mass to a change of force $\delta f$ is $1/m^2$. Since this holds both on the microscopic and macroscopic level,  the effective $m^2$ is the same as the microscopic one.

\subsection{Correlation lengths}

There are two correlations lengths for the interface. One is  
in the direction parallel to the interface $\xi_\parallel$, the other in the perpendicular direction $\xi_{\bot}$. They are both due to the confining potential, i.e.\ $m^2$,  but we name the parallel correlation length $\xi_m  = \xi_\parallel$, because it is the long-distance cut off set by $m$. We will place ourselves in the regime where the long-distance (infrared) cut off is not given by the system size $L$ but by $\xi_m$.
We define the \emph{roughness exponent} $\zeta$ as the exponent characterising the scaling of the lengths in the perpendicular direction with respect to the lengths in the parallel direction. At short distances  $u \sim x^\zeta$, see Fig.~\ref{fig:2ptTL92}, which translates into a relation between the two correlation lengths 
\be
\xi_{\bot} \sim \xi_m^\zeta.
\label{eq:roughness_xi}
\ee
The scaling properties of both qEW and qKPZ can be expressed as a function of $\xi_m$.

\subsubsection{Scaling of $\xi_m$ and $\xi_{\bot}$ for qEW}
In qEW the parameters $c$ and $m$ are protected by a symmetry, called the statistical tilt symmetry \cite{Wiese2021}. As a result, $c$ does not acquire an anomalous dimension under renormalization, thus  does not depend on $m$ and $\xi_m$.  
At depinning, all   forces scale in the same way, so equating the elastic force with the driving we get
\bea
 \nabla^2 u &\sim& m^2u  \nn  \\
\implies \quad \xi_m &\sim& m^{-1}.
\eea
From \Eq{eq:roughness_xi} we obtain
\be
\xi_{\bot} \sim m^{-\zeta}.
\ee

\subsubsection{Scaling of $\xi_m, \xi_{\bot}$ for qKPZ} \label{sec:corr_lenqkpz}
In qKPZ the term $\lambda (\nabla u)^2$ breaks the statistical tilt symmetry. As a consequence, $c$ is no longer protected, and acquires an anomalous dimension. Only $m$ is protected (see section \ref{sec:m_protection}). Assuming again that at depinning all   forces scale in the same way,
\bea
c\nabla^2 u &\sim& m^2u  \nn  \\
\implies  \quad  \xi_m &\sim& \frac{\sqrt{c}} m.
\label{eq:xi_m_scaling}
\eea
As we show below, 
$c$ increases when $m\rightarrow 0$. As a result,  the system has a larger correlation length than   in qEW.
Now that $\xi_m$ has a non-trivial scaling, we need a new exponent to describe this scaling. We chose to  use the scaling of $\xi_\bot$ with $m$, defining 
\be
\xi_{\bot} \sim m^{-\zeta_m}.
\ee
Using \Eq{eq:roughness_xi} we obtain
\be \label{eq:xim}
\xi_m \sim m^{-\frac{\zeta_m}{\zeta}}.
\ee

\subsection{Definition of the 2-point function}
\label{s:Definition of the 2-point function}

The 2-point function is defined as
\bea
\label{zeta-def}
C(x-y) &:=& \frac{1}{2} \overline{  [u(x){-}u(y)]^{2}} \nn\\
&\sim& \left\{\begin{array}{c}
A|x-y|^{2 \zeta},~~|x-y| \ll \xi_{m} \\
B m^{-2 \zeta_{m}}, ~~~~~~|x-y| \gg \xi_{m}
\end{array}\right..
\eea
The average is taken over   disorder configurations. 
We can formally define $\xi_\bot$ as 
\be\label{xi-perp-def}
\xi_\bot^2 := C(x-y)\Big|_{|x-y| \gg \xi_{m}}.
\ee
 $\xi_\parallel=\xi_m$  is the intersection point between the two asymptotic behaviors.
Taking $x= \xi_m$ in the 2-point   function, we get $A\xi_m^{2 \zeta} \simeq Bm^{-2\zeta_m}$, and as a consequence
\be
\xi_m  = C m^{-\frac{\zeta_m}{\zeta}}, \qquad C= \left(\frac{B}{A}\right)^{\frac{1}{2\zeta}}.
\ee
Let us stress that   key features of this universality class stem from $\frac{\zeta_m}{\zeta} \neq 1$.
This is illustrated with numerical results   for the  TL92 automaton  in $d=1$ in Fig.~\ref{fig:2ptTL92}, and for $d=2$ in Fig.\ \ref{fig:2ptTL922d}.  For anharmonic depinning  Figs.\ \ref{fig:2ptanh2d} and \ref{fig:2ptanh3d} show results in dimensions $d=2$ and $d=3$. Before we discuss them in depth, let us   extract the critical exponents in $d=1$ from directed percolation. This will serve as a strong check on our simulations.

\subsection{Connection to directed percolation}
\label{s:Connection to directed percolation} 
\begin{figure}
\centering
\setlength{\unitlength}{1cm}
{{\black
\begin{picture}(8.7,5.8)
\put(0,0){\includegraphics[width=8.4cm]{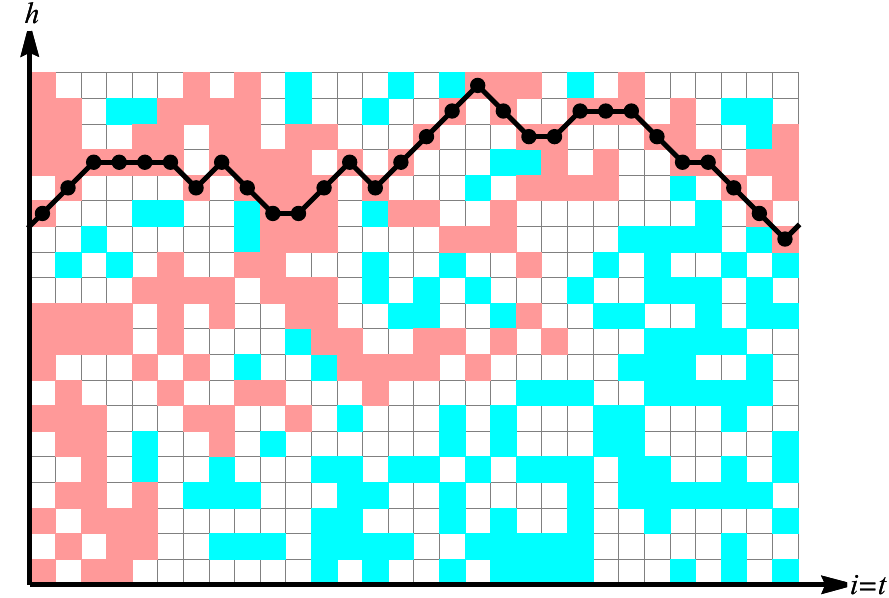}}
\put(7.7,0.35){$x,i=t$}
\put(0.25,5.55){$u$}
\end{picture}}}
\caption{Directed percolation from left to right. A site $(i,u)$  is  defined to be {\em connected} if it is occupied, and at least one of its left neighbors $(i-1,u)$,  $(i-1,u\pm 1)$ is connected.
The index $i$ takes the role of time $t$.}
\label{f:dir-percol2}
\end{figure}

In Sec. \ref{sec:mappings} we discussed that blocking configurations in TL92 are paths in directed percolation.
Here we extract the exponents  $\zeta$ and $\zeta_m$ from DP exponents. 
As the latter are  known precisely \cite{Hinrichsen2000,AraujoGrassbergerKahngSchrenkZiff2014,Dhar2017}, we get precise predictions for the former. 

There are two guiding principles for these relations: all forces at depinning have the same scaling dimension, and every length parallel to the interface   scales as $x$ or $\xi_m$, while lengths in the perpendicular direction   scale as $u \sim x^{\zeta}\sim m^{-\zeta_m}$.

Consider Fig.\ \ref{f:dir-percol2} which shows directed-percolation paths from left to right (in pink). 
They are constructed on a  square lattice, where occupied  cells (in pink or cyan) are selected with probability $p$, and the remaining once are unoccupied (white). A cell $(i,j) $ is said to be connected to the left boundary (and colored pink) if it is occupied, and if at least one of its three neighbors $(i-1, j)$ and $(i-1, j \pm1)$  is connected to the left boundary. The system is said to percolate if at least one point on the right boundary is connected to the   left boundary. To achieve periodic boundary conditions for TL92, it is further required that this remains true for the periodically continued system. 

While percolation is unlikely for small $p$, it is likely for large $p$, with a transition at $p=p_{\rm c}$.  
There are three independent  exponents $\beta$, $\nu_{\parallel}$, and $\nu_{\perp}$, defined via
\begin{align}
&\rho(t):=\left\langle\frac{1}{h} \sum_{u} s_{u}(t)\right\rangle \stackrel{t \rightarrow \infty}{\longrightarrow} \rho^{\text {stat}} \\
&\rho_{\text {stat}}  \sim\left(p-p_{\mathrm{c}}\right)^{\beta}, \quad p>p_{\mathrm{c}}, \\
&\xi_{\|} =\left|p-p_{c}\right|^{-\nu_{\|}}, \\
&\xi_{\perp}  =\left|p-p_{c}\right|^{-\nu_{\perp}}.  
\end{align}
Here $s_u(t)$ is the {\em activity} of site $u$ at time $t$, set to  one if the site is connected to the left boundary, and zero otherwise. $h=\sum_u$ is the height of the system, and  $\rho_{\text {stat}}$   the stationnary density of active sites. $\xi_{\|}  $ is  the size of the DP cluster  along the parallel  (time) direction,      and $\xi_{\perp} $ the size in the transverse direction. 
The last two relations imply
\be
\label{zeta-from-DP}
  \xi_{\perp} \sim \xi_{\parallel}^\zeta \implies \zeta^{d=1} = \frac{\nu_\perp}{\nu_\parallel} = 0.632613 (3).
\ee
This is the roughness exponent $\zeta$ defined in \Eq{zeta-def}. All numerical values are collected in table \ref{d=1-num-table}. 

For TL92, the surface  is blocked   by directed percolation paths in the direction parallel to the interface (from left to right).  However, instead of a global $p$, we have a $u$-dependent $p$, given by   $p-p_c  = m^2(u-w)$. As a result, the distance to  $p_{\rm c}$ in DP  corresponds   to a driving force   in  TL92.   Together with $u\simeq \xi_{\bot} \sim (p-p_c)^{-\nu_\bot}$, this gives $m^2 \sim (p-p_c)^{1+ \nu_\bot}$, or $(p-p_c)\sim m^{\frac{2}{1+ \nu_\bot}}$. This finally yields
\be
\label{zetam-from-DP}
u \sim m^{-\zeta_m} \implies \zeta_m^{d=1}  = \frac{2\nu_\bot}{1+ \nu_\bot} =  1.046190(4).
\ee
Note that in contrast to qEW   (where $\zeta_m= \zeta$), here  $\zeta_m > \zeta$.

 In $d\geq 2$  directed-percolation paths are
1-dimensional, whereas the interface is $d$-dimensional. As a result, the mapping breaks
down and one has to introduce directed surfaces
\cite{BarabasiGrinsteinMunoz1996}. Since no information for our simulations is gained, we will not discuss this case.

\begin{figure}
\centering
\includegraphics[width=\linewidth]{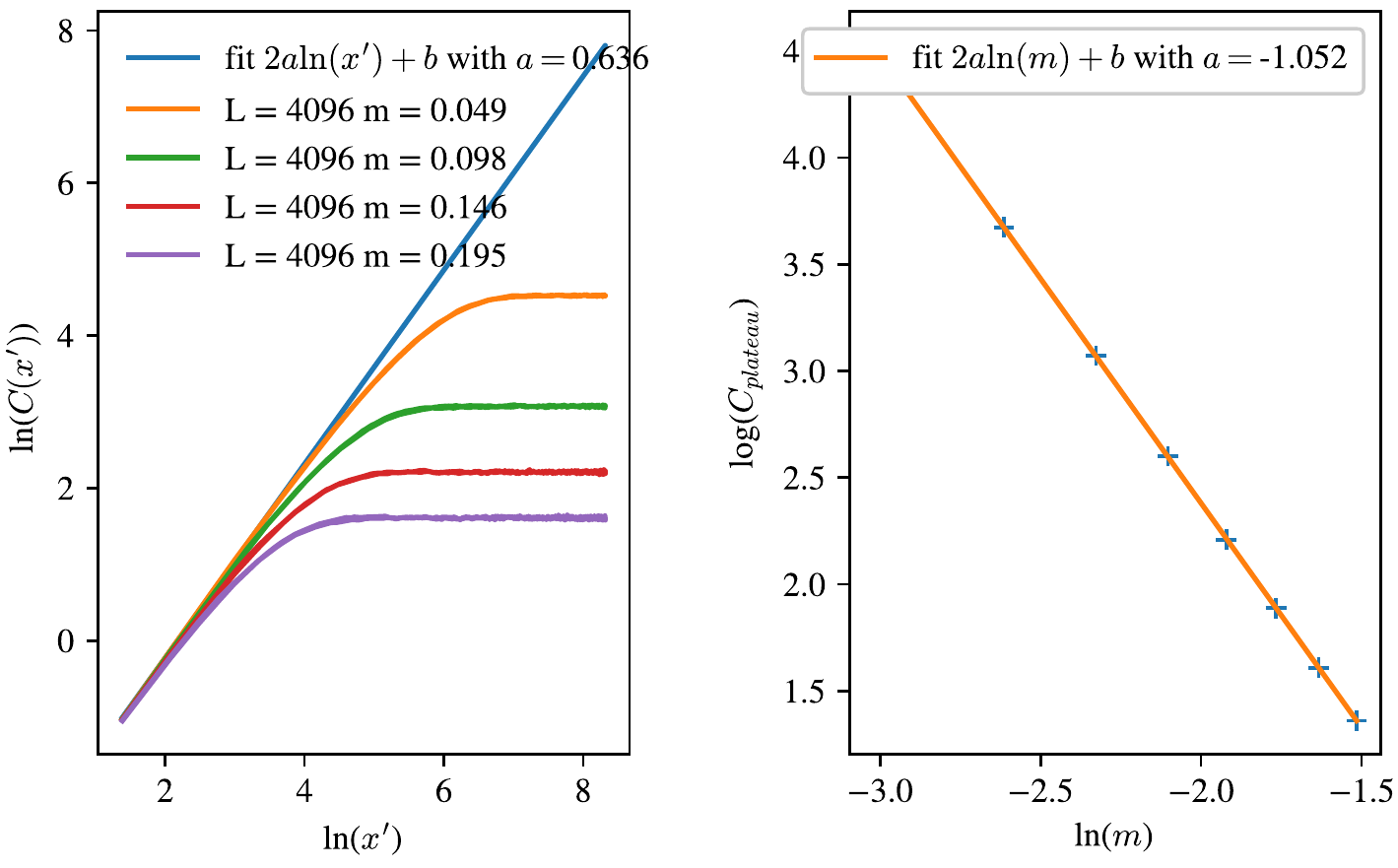}
\caption{TL92  $1d$ (left) 2-point function  $C(x)$  for different values of $m$ (not all shown here), plotted against $x^\prime = \frac{4x(L-x)}L $ to take advantage of the periodic boundary conditions. We read off the exponent $\zeta = 0.636$ in the linear part of the curve (the slope is $2\zeta$). (right) Scaling of the plateau of the 2-point functions for different $m$. The fit yields $\zeta_m = 1.052$.}
\label{fig:2ptTL92}
\end{figure}

\begin{figure}
\centering
\includegraphics[width=\linewidth]{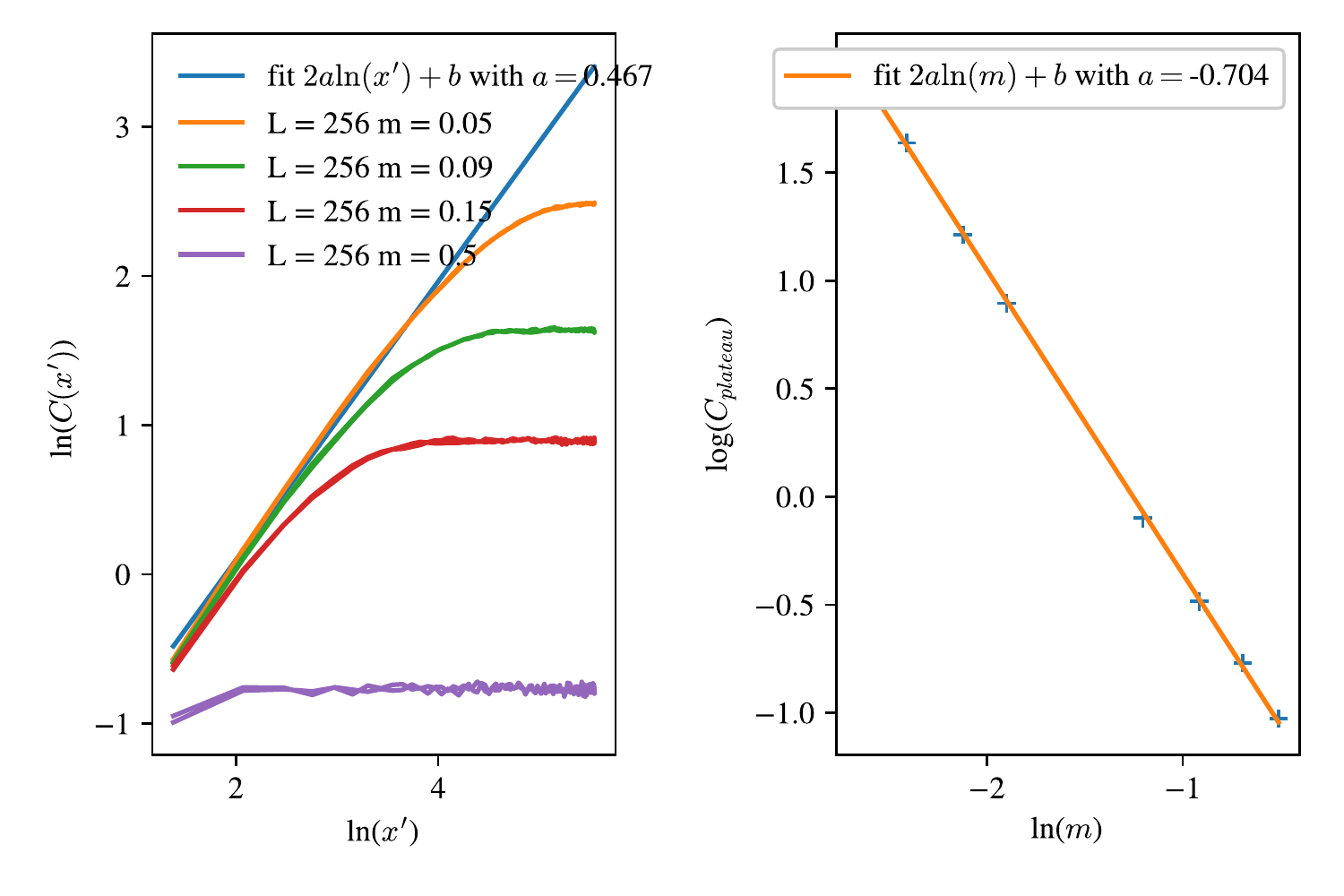}
\caption{ TL92  $2d$ (left) 2-point function $C(x)$ along  the diagonal of the system  for different values of $m$  (not all shown) plotted against $x^\prime = \frac{4x\sqrt2(L\sqrt2-x\sqrt 2)}{\sqrt2L} $. The exponent $\zeta \approx 0.47$ is obtained from  the linear part of the curve. (right) Scaling of the plateau of the 2-point functions for different $m$. The fit yields $\zeta_m = 0.70$.}
\label{fig:2ptTL922d}
\end{figure}

\begin{figure}
\centering
\includegraphics[width=1\linewidth]{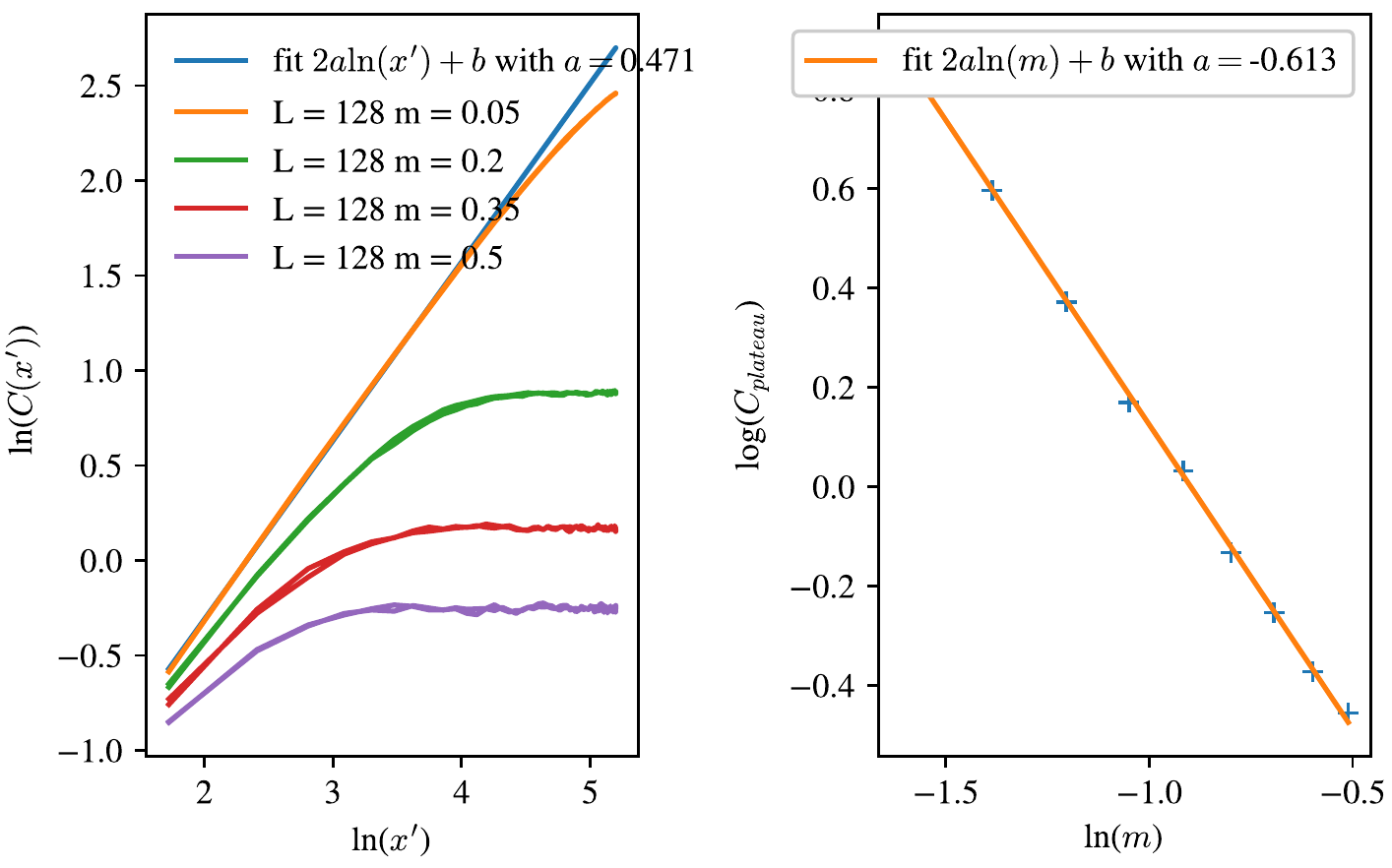}
\caption{Anharmonic depinning $2d$ (left) 2-point function $C(x)$  alongside the diagonal of the system  for different values of $m$  (not all shown here) plotted against $x^\prime = \frac{4x\sqrt2(L\sqrt2-x\sqrt 2)}{\sqrt2L} $. The exponent $\zeta = 0.47$ is extracted from the linear part of the curve. (right) Scaling of the plateau of the 2-point functions for different $m$. The fit gives $\zeta_m = 0.61$.}
\label{fig:2ptanh2d}
\end{figure}

\begin{figure}
\includegraphics[width=1\linewidth]{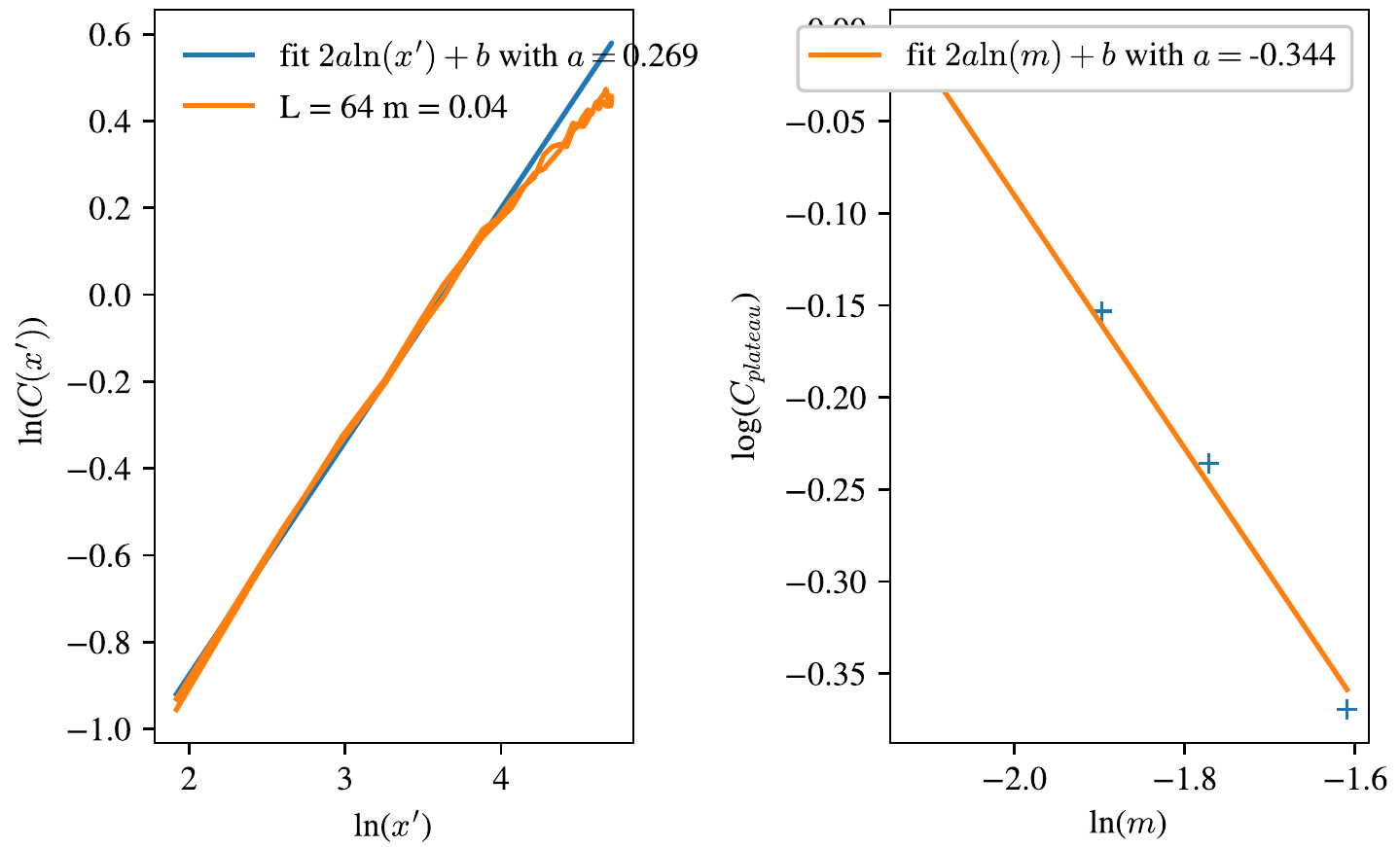}
\caption{Anharmonic depinning $3d$ (left) 2-point function $C(x)$ alongside the diagonal of the system   for different values of $m$  (not all shown) plotted against $x^\prime = \frac{4x\sqrt3(L\sqrt3-x\sqrt 3)}{\sqrt3L} $. The exponent $\zeta = 0.27$ is read off from  the linear part of the curve. (right) Scaling of the plateau of the 2-point functions for different $m$. The fit gives $\zeta_m = 0.34$.}
\label{fig:2ptanh3d}
\end{figure}

\subsection{Results for the 2-point function, $\zeta$ and $\zeta_m$}

For  TL92  in $d=1$, the 2-point function  is shown on Fig.~\ref{fig:2ptTL92}.  
$d=2$ is covered in Figs.~\ref{fig:2ptTL922d}-\ref{fig:2ptanh2d}, while Fig.~\ref{fig:2ptanh3d} is for dimension $d=3$.
The results for the critical exponents  $\zeta$ and $\zeta_m$ are summarized in Table \ref{tab:flow}. 

Let us first discuss our choice of simulation parameters. 
To obtain      $\zeta$, the smallest possible $m$ is chosen, such that there is no system-spanning avalanche. 
The latter would  mix the physics of the $d$-dimensional interface with that of a single degree of freedom. For $\zeta_m$ one needs a value of $m$ that allows to  clearly see the plateau of the 2-point function. 
Finally, we saw with seemingly   little noise the 2-point function for larger systems, until $L=1024$ for  TL92,  but we found systematic errors in those bigger systems, due to a lack of  statistics. As a  rule of thumb,  $2\times 10^4$ avalanches/per site are necessary to ensure a scaling collapse of the 2-point function for different sizes.

Let us now discuss our results, summarized on tables \ref{tab:flow} and \ref{d=1-num-table}.
In $d=1$, there are consistent values for $\zeta$ and $\zeta_m$ between the three simulated models, and directed percolation. We thus have confirmed numerically that there is a single universality class, and that the scaling arguments for $\zeta$ (known in the literature) and  $\zeta_m$ (introduced here) are valid.

In $d=2$, our simulations show that TL92   and anharmonic depinning share the same exponents.
This is consistent with the mapping established in section \ref{Cellular automata in arbitrary dimension}.

In  $d=3$, the exponents seemingly   differ, suggesting that the two universality classes   may be  different. 
This is consistent with the absence of a mapping established at the end of  section \ref{Cellular automata in arbitrary dimension}. 
On the other hand, we cannot exclude that  finite-size corrections, which are expected to be large for a cellular automaton such as TL92, are responsible for this lack of agreement.

\begin{table}[t]
\begin{ruledtabular}
\begin{tabular}{@{}llllll@{}}
\textrm{model}&
\textrm{$\zeta$}&
\textrm{$\zeta_m$}&
\textrm{$\zeta$ literature}\\
\colrule
aDep 1d &  $0.635(6)$ & $1.054(3) $ & $0.63$  \cite{RossoHartmannKrauth2002} \\
TL92 1d & $0.636(8)$ & $1.052(5) $& $0.63$ \cite{AmaralBarabasiBuldyrevHarringtonHavlinSadr-LahijanyStanley1995}  \\
qKPZ 1d & $0.64(2)$ & $1.05(1) $& $0.633 (8) $ \cite{LeeKim2005} \\
\hline
qEW 1d & $1.25(1)$ & $1.25(1)$ &  $1.25$ \cite{RossoHartmannKrauth2002,FerreroBustingorryKolton2012,GrassbergerDharMohanty2016}\\
\hline\hline
aDep 2d & $0.48(2) $ &  $0.61(2) $  & $0.45 (1) $\cite{RossoHartmannKrauth2002} \\
TL92 2d & $0.47(3)$ & 0.70(3) & $0.48 (3)$ \cite{AmaralBarabasiBuldyrevHarringtonHavlinSadr-LahijanyStanley1995} \\
\hline
qEW 2d & - & - & $0.753 (2)$  \cite{RossoHartmannKrauth2002} \\
\hline\hline
TL92 3d & $0.44(5)$ & $0.52(6)$ & $0.38 (4)$ \cite{AmaralBarabasiBuldyrevHarringtonHavlinSadr-LahijanyStanley1995}  \\
aDep 3d & $0.27(4)$ & $0.34(4)$ & $0.25 (2)$ \cite{RossoHartmannKrauth2002} \\
\hline
qEW 3d & - & - & $0.355 (1)$  \cite{RossoHartmannKrauth2002} \\
\end{tabular}
 \caption{Critical exponents $\zeta$ and $\zeta_m$ of the qEW and qKPZ classes, from simulations of anharmonic depinning and comparison with the literature.  Only $\zeta$ was measured in the literature, while $\zeta_m$ is also necessary to describe the qKPZ class (see Section \ref{sec:corr_lenqkpz}).}
 \label{tab:flow}
\end{ruledtabular}
\end{table}

\begin{table}[t]\begin{ruledtabular}
\begin{tabular}{@{~~~}ccc  @{~~~}}
\textrm{exponent}&
\textrm{DP value}&
\textrm{simulated value}\\
\colrule
$\nu_{\parallel}$&$1.733847(6)$, &\\
$\nu_\bot$ & $1.096854(4)$,&\\
\\
$\zeta$& $0.632613(3)$,& $0.636(4)$\\
$\zeta_m$&  $1.046190(4)$,&$1.052(6)$ \\
  $\frac{\zeta_m}{\zeta}$& $1.65376(1)$,  &$1.65(1)$ \\
$\tau$& $1.259246(3)$,& $1.257(5)$\\
$\psi_{c}$& $1.30752(2)$, & $1.31(4)$\\
$\psi_{\lambda}$& $0.26133(2)$,& $0.28(3)$\\
\\
$z$&-&$1.10(2)$\\
$\alpha$&-&$1.28(3)$\\
$\psi_\eta$ &-& $-0.18(1)$\\
$\beta$ &-& $0.81(3)$\\
\end{tabular}
 \caption{The DP   values for all exponents are  from Ref.~\cite{Hinrichsen2000} (first two lines),  combined with the  scaling relations derived here (following lines). The agreement between the static exponents   numerically estimated   for aDep and TL92 and their DP  values is excellent. There is no such mapping for dynamical exponents. The conjecture $z=1$ advanced in Ref.~\cite{AmaralBarabasiBuldyrevHarringtonHavlinSadr-LahijanyStanley1995}  is in contradiction to our simulations, see Appendix \ref{s:zconjecture} for a detailed discussion.}
\label{d=1-num-table}
\end{ruledtabular}
\end{table}

\subsection{The exponent $\nu$}
By definition of the correlation-length exponent $\nu_{\parallel}$,
$ \xi_m \sim (f-f_c)^{-\nu_{\parallel}}$, with $f$ the driving force, and $f_c$ the critical depinning force. 
This identifies the standard depinning exponent $\nu$ as 
\be
\nu^{d=1} = \nu_\parallel.
\ee
Since $(f-f_c)  = m^2(u-w) \sim m^{2-\zeta_m}$, together with \Eq{eq:xim} this implies  
\be
\frac{\zeta_m}{\zeta} = \nu (2-\zeta_m). 
\label{eq:nupar}
\ee
This relation is valid in any dimension $d\leq d_c$ and does not rely on the mapping to DP.
In  $d=1$ replacing $\zeta_m$ and
$\zeta$ by their expressions  in terms of $\nu_{\bot}$ and
$\nu_{\parallel}$ given in   Sec.~\ref{s:Connection to directed percolation}, we   verify consistency.

\subsection{Dynamical exponent $z$}
The response function $R(x,t)$ is   defined as the response of the system at time $t$ and position $x$, given a kick in the force of the confining potential $m^2 w(x,t)$ at time $t=0$ and $x=0$ (we use translational invariance in both space and time),
\be
R(x,t) :=  \frac{\delta \left< u(x,t)\right>}{ m^2 \delta w(0,0)} .
\ee
Assume that the response function takes the scaling form
\be
R(x,t)  = \frac{1}{m^2t^{\frac d z}} f\Big(\frac{x}{t^{\frac1z}} \Big),
\ee
where $\int_x f(x)=1$. Then $z$ is the dynamical critical exponent.

For 
the velocity of an avalanche by
definition  $v \sim (f-f_c)^\beta$, and $v = \frac{u}{t} = \frac{\xi_{\bot}}{\xi_{\parallel}^{z}} =\xi_{\parallel}^{\zeta - z} = (f-f_c)^{-\nu_\parallel (\zeta -z)}   $. Eliminating  $\nu_{\parallel}$ with the help of \Eq{eq:nupar} we get
\be
\beta = \frac{\zeta_m(z-\zeta)}{\zeta(2-\zeta_m)}.
\label{eq:beta}
\ee
Scaling relations for qEW  are recovered by setting $\zeta_m = \zeta  $, resulting in 
\be
\beta_{\rm qEW} = \frac{z-\zeta}{2-\zeta}= \nu (z-\zeta).
\ee
We  evaluated  $z$ in the TL92 automaton in $d=1$ by looking at the joint distribution of avalanche duration $T$ and lateral extension $\ell$, shown in Fig.~\ref{fig:avalanchejointdistrib} (see section \ref{sec:avduradistrib} for  details). 
Using 
\be\label{39}
T \sim \ell ^z, 
\ee
we find  
\be
z^{d=1}_{\rm TL92}=1.10 \pm 0.02.
\ee 
This value contradicts   Ref.~\cite{AmaralBarabasiBuldyrevHarringtonHavlinSadr-LahijanyStanley1995}, which conjectures the \emph{exact} value $z=1$ using heuristic arguments and   evidence   from numerical simulations. Our simulation, like theirs, computes the lateral extension of an avalanche as a function of its duration. While Ref.~\cite{AmaralBarabasiBuldyrevHarringtonHavlinSadr-LahijanyStanley1995}  extracts the power law by a fit to one decade,  we have data on more than $2.5$ decades, allowing for a much more precise value. 
We reviewed the   argument given in Ref.~\cite{AmaralBarabasiBuldyrevHarringtonHavlinSadr-LahijanyStanley1995}, which relies on shortest paths   on a percolation cluster. Our main criticism is that 
    transport properties on percolation clusters are linked to the proportion of singly-connected bonds, bonds that if cut, separate the percolation cluster into two parts, (the ``blobs and links" representation \cite{StaufferAharony1994}). Even if the perpendicular direction is small compared to the longitudinal one, it is non zero, which  changes   the proportion of singly connected bonds. Details are given in Appendix \ref{s:zconjecture}.

\begin{figure}[t]
\centering
\includegraphics[width=1\linewidth]{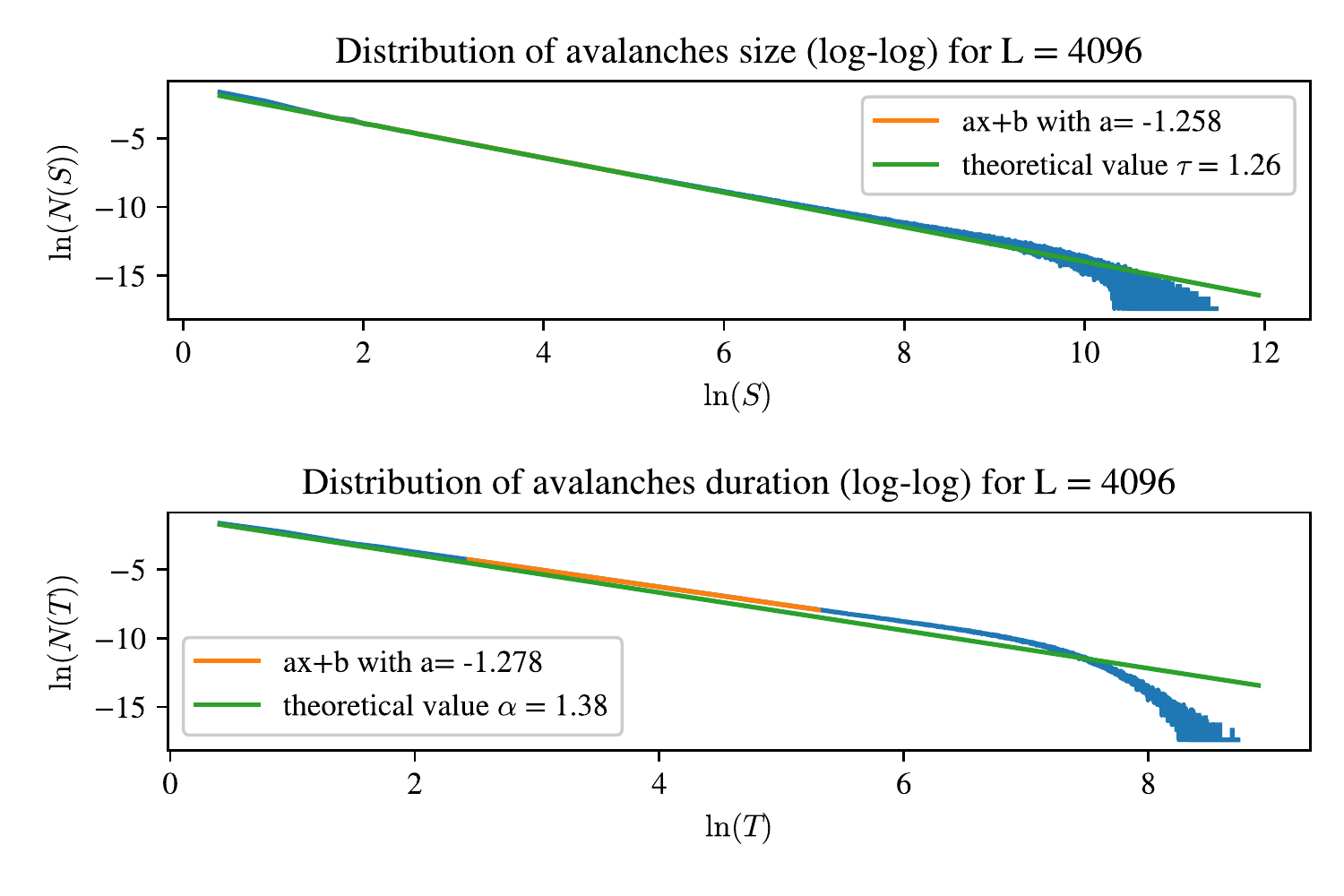}
\caption{Avalanches-size and duration distributions for TL92. $\tau$ and $\alpha$ are the associated exponents. We also computed $S_m$ and $T_m$ from these distributions at different $m$ and verified the relations \Eq{eq:sm} and \Eq{eq:tm} (not shown). } 
\label{fig:avalancheexp}
\end{figure}
\subsection{Avalanche size}
Let us recall   scaling   for   avalanches, adapted to   qKPZ. Let 
$S$  the size of an avalanche, i.e.\ is the  number of sites that are affected in an avalanche (in a cellular automaton), or the volume swept through by the interface between two blocking configurations. 
Define its typical size $S_m$ as \cite{LeDoussalMiddletonWiese2008}
\be
S_m:= \frac{\langle S^2\rangle}{2 \langle S\rangle}.
\label{eq:s_m}
\ee
If the avalanche-size distribution decays as an exponential for large $S$, then this exponential decay is $\sim \rme^{-S/(4S_m)}$ \cite{LeDoussalMiddletonWiese2008,ChenPapanikolaouSethnaZapperiDurin2011,KoltonLeDoussalWiese2019}, identifying $S_m$ as the large-scale cutoff. Note that using \Eq{eq:s_m}  is very precise,  while fitting a tail is rather imprecise.

Scaling implies that 
\be
\label{41}
  S \sim \ell^{d+\zeta}\quad \implies \quad  S_m \sim \xi_{m}^{d + \zeta},
\ee 
where $\ell$ is the lateral extension of an avalanche. 
Injecting \Eq{eq:xim} yields 
\be
  S_m \sim m^{-d \frac{\zeta_m}{\zeta}- \zeta_m}.
 \label{eq:sm}
\ee
Assume  that  $ P_S(S) \sim S^{-\tau} $ for $S \ll S_m$.
To obtain a scaling relation for $\tau$, follow \cite{DobrinevskiLeDoussalWiese2014a} to consider  the avalanche-size distribution per unit   force,
\be
\rho_{f}(S):=\frac{\delta N(S)}{\delta f} \simeq S^{-\tau} f_{S}\left(S / S_{m}\right) g_{S}\left(S / S_{0}\right), ~~ S_{0} \ll S_{m}.
\label{34}
\ee
$S_m$ is the large-scale cutoff introduced above,  while $S_0$ is a small-scale cutoff. We expect \Eq{34} to have a finite limit  when $m \rightarrow 0$ \cite{DobrinevskiLeDoussalWiese2014a}. Associated to a force  increase by $\delta f$ is a  total displacement $\int_x \delta u(x) = \langle S\rangle$. The total increase in force can be written as $ \delta f = \int_x m^2 \delta u(x) = m^2 \langle S\rangle$.
By the definition of $\rho_f$ we have $\langle S\rangle = \delta f \int^\infty_0 S  \rho_f(S)   \rmd S$. This gives
$$
1 = m^2 \left[ S_m ^{2-\tau} - \ca O(S_0 ^{2-\tau}) \right].
$$
Since $\tau <2$, we can take the limit of $S_0\to 0$, resulting in
\be
 \tau=2-\frac{2}{d \frac{\zeta_{m}}{\zeta}+\zeta_{m}}.
\ee
We compare this result to simulations in section \ref{Numerical simultions for size and duration} below.

\begin{figure*}[t]
 \includegraphics[width=0.9\textwidth]{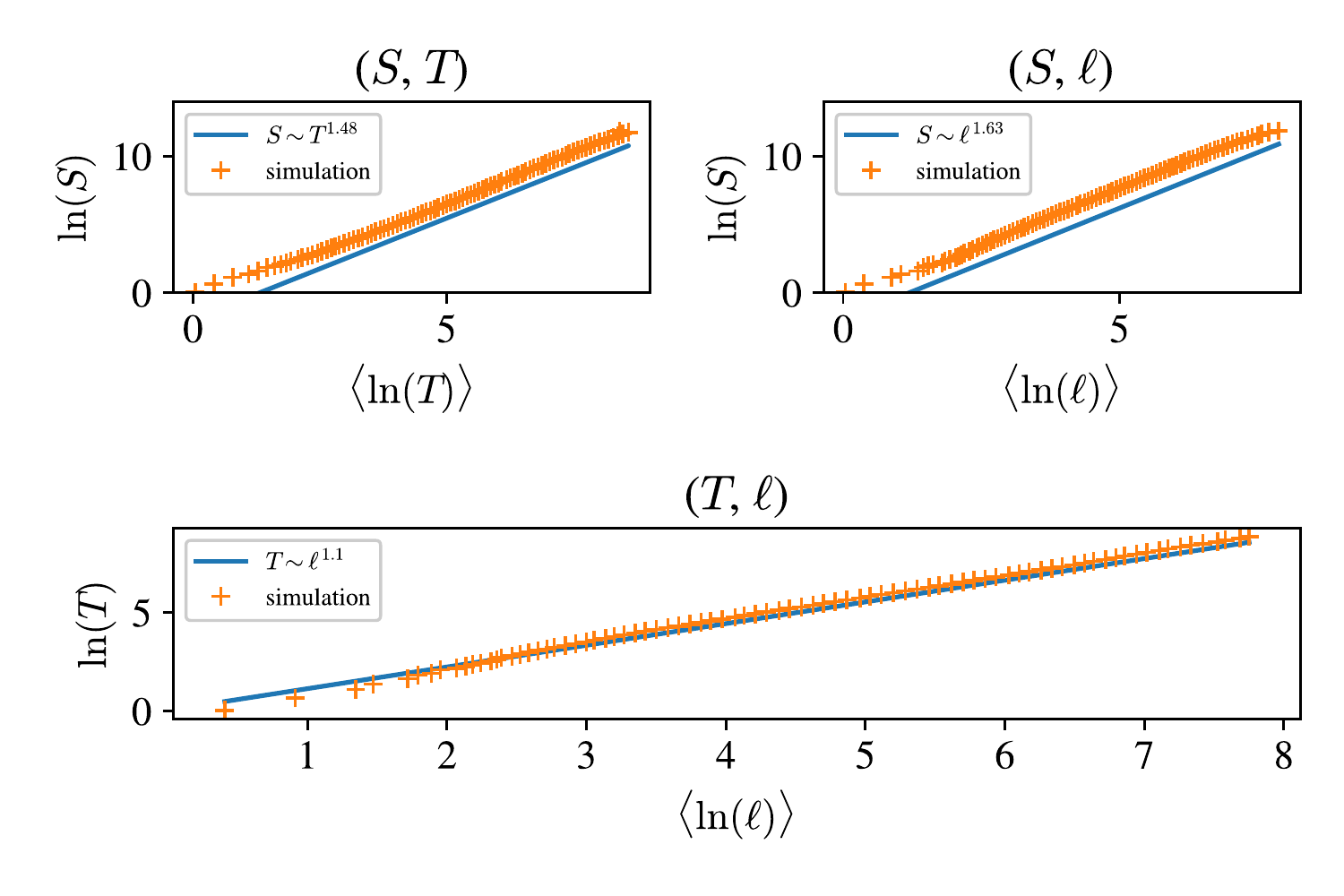}
\caption{Joint probability distributions of $(S, \langle T\rangle), (S, \langle\ell\rangle), (T, \langle\ell\rangle)$ in TL92, in log scale, for $d=1$,  for $L=4096$ and $m=0.024$. They verify   \Eqs{eq:st},  \eq{41} and the definition of $z$ in \Eq{39}.  Averaging before or after taking the log gives similar results. The solid line is the theoretical prediction.}
\label{fig:avalanchejointdistrib}
\end{figure*}

\subsection{Avalanche duration}\label{sec:avduradistrib}

Consider the dynamics of an avalanche, with $\ell$ its lateral extension and  $T$ its duration. 
Using $T \sim \ell ^z$ and $S\sim \ell^{d+\zeta}$, we get
\be
S \sim T ^{\frac{d+ \zeta}{z}}.
\label{eq:st}
\ee
Assume that 
\be
P_T(T) \sim T^{-\alpha} \quad \mbox{for}~ T\ll T_m:= \frac{\left<T^2\right>}{2\left<T\right>}.
\label{eq:tm}\ee
Scaling implies that $P_S(S)\rmd S \sim P_T(T) \rmd T $. For small avalanches  (but  bigger than the discretization cutoff) this implies that $S^{1-\tau } \sim T^{1-\alpha}$.
Using \Eq{eq:st} we obtain 
\be
\alpha = 1 + \frac1z \left(  {d+\zeta}  - \frac{2\zeta}{  \zeta_m}\right).
\label{eq:scalingalpha}
\ee

\subsection{Numerical simulations for size and duration}
\label{Numerical simultions for size and duration}

Let us first explain our choice of parameters:
 to study avalanches, it is important to avoid triggering two overlapping avalanches; to that end, we use a driving strength $w \rightarrow  w + \delta w$  such that the probability that a site gets depinned is $\frac{1}{40}$. As a comparison, in the other simulations we generate on average one avalanche per driving event.
Next, $m$ should be large enough to avoid system-spanning avalanches. For $L = 4096$ we computed our distributions for $ 2\times 10^7$ avalanches and $m = 0.0244$.

We verified the  scaling relations for the dynamic exponent $z$ and
the size and duration exponents $\tau$ and $\alpha$. 
To this end, we recorded for   TL92  in $d=1$ the joint distribution of
$(S, T, \ell)$, with $\ell$ the lateral extension of the avalanche.
This allowed us to extract three joint distributions involving two variables, and shown on  Fig.~\ref{fig:avalanchejointdistrib}.
First, we use $T\sim \ell^z$ to extract $z$ in $d=1$ as 
\be
z^{d=1}= 1.10(2).
\ee
This gives for the remaining relations the numerical values 
$
T\sim \ell^{1.48(3)}$, and  $S\sim \ell^{1.6326(3)}$. A glance on Fig.~\ref{fig:avalanchejointdistrib} shows that the data are in good agreement with these values. 

Fig.~\ref{fig:avalancheexp} shows the size and duration distributions, with predicted exponents $\tau=1.2592(6)$, and $\alpha=1.385(7)$.
While the former is satisfied over almost three decades, the latter seemingly comes out much smaller, namely at 
\be
\alpha_{\rm TL92}^{d=1}= 1.28(2) 
\ee 
Let us discuss possible sources for this discrepancy:

\noindent{(i)}
 The real functional form of $P_T$ is more complicated than the scaling anzatz in \Eq{eq:tm}, and has a ``shoulder" that   pushes the apparent exponent down. This phenomenon was described   for the size distribution  $P_S$ in qEW,  both numerically \cite{RossoLeDoussalWiese2009a} and within the FRG \cite{LeDoussalWiese2008c}; it was studied numerically for $P_S$ on qKPZ   \cite{ChenPapanikolaouSethnaZapperiDurin2011}. As the top plot of Fig.~\ref{fig:avalancheexp} shows, there is a small shoulder for $P_S$, but the agreement on the exponent is very good. If the shoulder for $P_T$ is   much longer, it is hard to see on Fig.~\ref{fig:avalancheexp}.

\noindent{(ii)} We still see large finite-size corrections due to the discretization of the time evolution. This would be surprising in view of the excellent scaling in the $(T,\ell)$ and $(S,T)$ plots of Fig.~\ref{fig:avalanchejointdistrib}.
 
We could properly simulate avalanche durations  only in a cellular automaton, since for anharmonic depinning we used the variant Monte Carlo algorithm of  \cite{RossoKrauth2001a,RossoHartmannKrauth2002}, which has a different (probably faster) time evolution. Whether this amounts to a smaller exponent $z$ is however doubtful.

\subsection{Comparison with qEW}
What is the effect of   the non-linearity on the physics of the system? Can one get an intuition?
The increase of the short-range elasticity with the scale has two main effects: the roughness exponent $\zeta$ decreases     from $1.25$ for qEW  to $\zeta = 0.63$ for qKPZ, meaning  the width is reduced at large scales. 
The parallel correlation length $\xi_m$ for $m\rightarrow 0$  grows faster than for qEW, reflected in $\frac{\zeta_m}{\zeta} > 1$. As the elasticity at large scales gets stronger, more sites are correlated  and the correlation length  increases.
 The avalanche size exponent $\tau$ goes from $\tau_{\text{qEW}} = 1.11$ \cite{AragonKoltonDoussalWieseJagla2016} in $d=1$, to $\tau = 1.26$, close to the one in dimension $d=2$ for qEW. 

 \begin{table}[h]
\begin{tabular}{  c @{\qquad} c }
\toprule
\textrm{qEW}&
\textrm{qKPZ}\\
\colrule
$\xi_m \sim m^{-1} $& $\xi_m \sim m^{-\frac{\zeta_{m}}{\zeta}}$\\
\\
$S_m \sim m^{-d-\zeta} $ & $S_m \sim m^{-d\frac{\zeta_{m}}{\zeta}-\zeta_m} $ \\
\\
$\tau=2-\frac{2}{d +\zeta}$ &$\tau=2-\frac{2}{d \frac{\zeta_{m}}{\zeta}+\zeta_{m}}$\\\botrule
\end{tabular}
\caption{\label{tab:comp_qewqkpz}%
Scaling relations for qEW can be obtained from qKPZ by setting $\zeta_m = \zeta$. To pass from qEW to qKPZ, it suffices to replace $d$ by $d \frac{\zeta_{m}}{\zeta}$ and $\zeta$ by $\zeta_m$ when it is linked to a length in the $u$ direction.
}
\end{table}

\section{ Effective force correlator $\Delta(w)$ and running coupling constants}
\label{s:Effective action}
\subsection{Definition of the effective force correlator $\Delta(w)$}

 In \Eqs{qEW}, \eq{eq:qkpz} and \eq{eq:anhdep} we had introduced a restoring force  $m^2[w-u(x,t)]$ from a confining potential. This was not only necessary to drive the system, but also to  estimate  the  effective force correlator $\Delta(w)$, by measuring the fluctuations of the center-of-mass position $u_w$ of the interface.  Define $\Delta(w)$   as
\bea\label{Delta-def}
\Delta(w-w') &:=& m^4 L^d\, \overline{ (u_w-w) (u_{w'}-w') }^{\rm c}, \\
u_w &:=& \frac1{L^d} \int_x u_w(x), \\
u_w(x) &:=& \lim _{t\to \infty} u(x,t) \mbox{~given~}w\mbox{~fixed.}
\label{uw}
\eea
In   our protocol, $w$ is increased in steps. One then waits until the interface stops, which defines $u_w(x)$. Its center-of-mass position is $u_w$, and its fluctuations define $\Delta(w)$.

\subsection{Scaling of  $\Delta(w)$}\label{sec:scalingdelta}
The definition \eq{Delta-def} has a finite limit for fixed $m$, when $L\to \infty$.
Using that $u\sim m^{-\zeta_m}$, and that $L/\xi_m$ is dimensionless leads together with \Eq{eq:xim} to
\be
\Delta(0) \sim   m^{4-d \frac{\zeta_{m}}{\zeta}-2 \zeta_{m}}.
\label{fig:scalingdelta0}
\ee
For the argument of $\Delta(w)$, we expect 
\be\label{w-u-scaling}
w \sim u \sim m^{-\zeta_m}.
\ee
A non-trivial check is as follows: 
As in qEW, 
one can connect the typical avalanche size given in \Eq{eq:s_m} to the disorder force correlator (see e.g.~\cite{Wiese2021}, Eq.~(104)),
\be
  |\Delta^\prime(0^+)| = m^4 \frac{\langle S^2\rangle}{2 \langle S\rangle} \sim m^{4-\zeta_m(d/\zeta +1)}.
  \label{eq:DeltaSm}
\ee
This is consistent with \Eqs{fig:scalingdelta0} and \eq{w-u-scaling}.

\subsection{Measuring $\Delta(w)$}
\label{s:Measuring Delta}
$\Delta(w)$ is defined from the  variable $u_w$ in \Eq{uw}. For depinning, be it anharmonic or not, integrating the equation of motion with periodic boundary conditions leads to 
\bea
m^2(u_w-w) &=& F_w,\\
F_w &:=& \frac1{L^d} \int_x F(x,u_w(x)).
\eea
Thus $\Delta(w)$ is also the correlator of the disorder acting on the interface. This direct connection breaks down in qKPZ, as after integration over the center of mass three terms remain: $m^2 (w-u_w)$, $F_w$, and $\Lambda_w$, defined by 
\be
\Lambda_w:= \frac 1{L^d}\int_x \lambda [\nabla u_w(x)]^2.
\ee 
A configuration at rest then has 
\be
m^2 (w-u_w)+ F_w+ \Lambda_w = 0.
\ee
We note that while the first and last term are positive, the middle term is negative. So why did we define  $\Delta(w)$ as the (connected) correlations of $u_w$, and not $F_w$? After all, we call it the {\em renormalized force correlator}. The answer comes from 
 more sophisticated field theory arguments, developed in a companion paper \cite{MukerjeeWiese2022}.
In essence it looks at all 2-time contributions to the $uu$ correlations, and then amputates the external response functions. The result is as given in \Eq{Delta-def}. For details we refer to \us.

We have verified that \Eqs{fig:scalingdelta0}-\eq{eq:DeltaSm} hold for  TL92,  and the other two models.
 The correct regime to obtain a good scaling collapse for the correlator -- with the exponents given in Table \ref{tab:flow} --  is when the infrared cutoff is set by the confining parabola, meaning that the plateau of the 2-point function is reached. 
The results for the shape of $\Delta(w)$ are summarized in Figs.\ \ref{fig:cors1} and  \ref{fig:cors2}, where everything is rescaled such that  $|\Delta^{\prime}(0^{+})| = \Delta(0) = 1 $.

\begin{figure}
\!\!\!\includegraphics[width=1.0\linewidth]{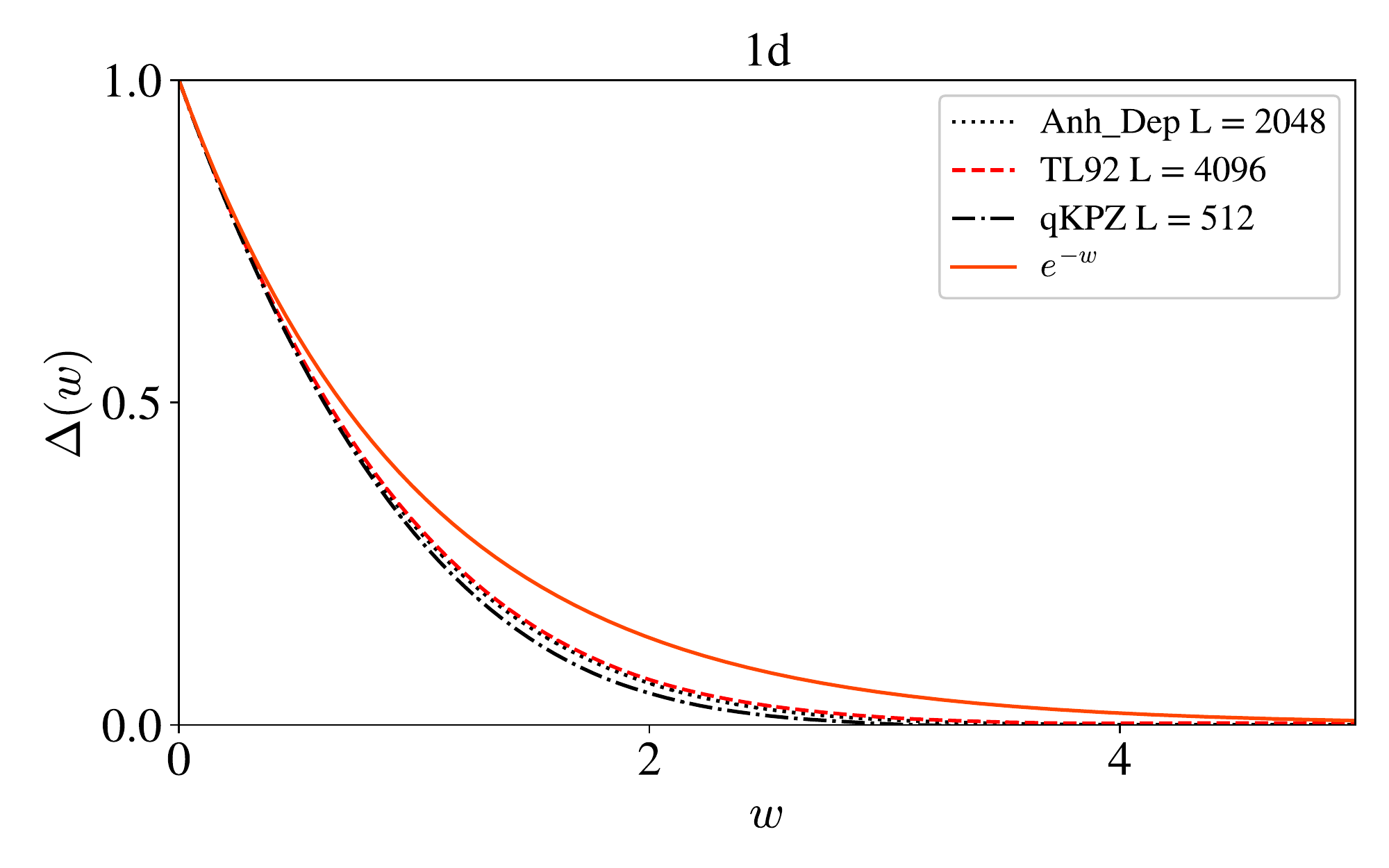}
\caption{Comparison of the shapes of the correlators in $d=1$ and for anharmonic depinning,  TL92   and qKPZ  with the exponential behavior observed in a subspace  in Ref.~\cite{LeDoussalWiese2002}. We    clearly see (a): The correlators for TL92, aDep and qKPZ are very close.  (b) The  subspace found in \cite{LeDoussalWiese2002} is not the one attained by the evolution of those models. The shapes have been normalized by setting $|\Delta^{\prime}(0^{+}) |= \Delta(0) = 1 $.}
\label{fig:cors1}
\end{figure}

\begin{figure}
\Fig{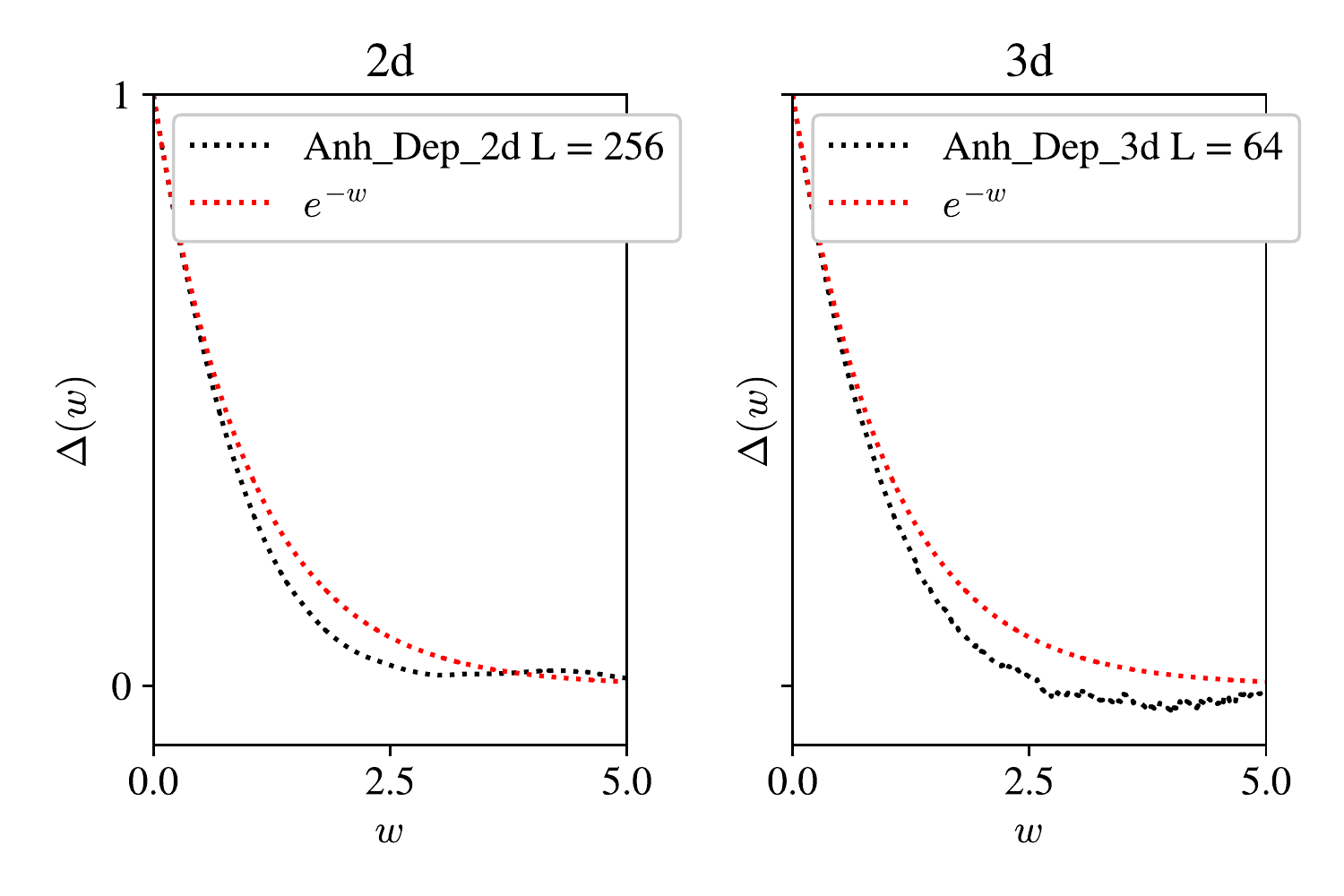}
\caption{Comparison of the shapes of the correlators in $d=2,3$   for anharmonic depinning with the exponential behavior found for a subspace in Ref.~\cite{LeDoussalWiese2002}. We    see  that  this subspace  is not the one attained by   these models. The shapes are   normalized s.t.\ $|\Delta^{\prime}(0^{+}) |= \Delta(0) = 1 $.}
\label{fig:cors2}
\end{figure}

\subsection{Anomalous dimensions for $c$, $\lambda$, and  $\eta$}
If there were no corrections to $c$, $\lambda$ and $\eta$, the theory would be trivial. Before we show in the next section \ref{sec:algorithm} an algorithm to   estimate   their scale dependence, let us first derive their anomalous dimensions, given the information already obtained.

Let us define their scaling dimensions   as  
\bea 
\lambda &\sim&  m^{-\psi_\lambda}\\
c &\sim& m^{-\psi_{c}}\\
\eta &\sim& m^{-\psi_{\eta}}.
\eea 
Equating the dimensions of   driving force and   elasticity,   $c \nabla^2  u \sim m^2 u  $, we get
$m^{-\psi_{c}-\zeta_m}   \xi_m^{-2} \sim m^{2-\zeta_m} $ and together with \Eq{eq:xim}
\be
\psi_{c} = 2 \frac{\zeta_m - \zeta}{\zeta}.
\label{eq:scalingc2}
\ee 
A similar argument for $\lambda$ yields
\be
\psi_\lambda = 2 \frac{\zeta_m - \zeta}{\zeta} - \zeta_m.
\label{eq:scalinglamb}
\ee
These two relations have been verified (see left of Fig.\ \ref{fig:fixedpoint}),  thanks to the algorithm we describe in Sec.~\ref{sec:algorithm}.

The scaling relation for $\psi_{\eta}$ is obtained from $\eta \partial_t u \sim m^2 u $, implying $t\sim m^{-2-\psi_\eta}\sim x^{(2+\psi_\eta)\zeta/\zeta_m}$. This yields 
\be
 \psi_\eta  = z\frac{\zeta_m}{\zeta} - 2.
\ee

\subsection{An algorithm to   estimate   the effective coupling constants}\label{sec:algorithm}

In order to obtain the effective KPZ non-linearity $\lambda$, one can tilt the system and   estimate  the change in the depinning force as in \cite{TangKardarDhar1995}. In contrast, the effective elasticity $c$ has to our best knowledge never been   estimated numerically. 
Since   the field theory in Ref.~\cite{LeDoussalWiese2002} did not deliver an  FRG fixed point for the ratio  $\lambda/c$, we decided to check numerically whether such a fixed point exists, and to extract as much information as possible to constrain the field theory.

Our algorithm to achieve this  is simple: measure the response of the interface to a perturbation, sinusoidal in space, and constant in time.
This is achieved by driving the system with a spatially modulated background field $w(x)$, see Fig.\ \ref{fig:sinus_driving}, 
\be\label{w(x)}
w(x) = w_0 + A \sin \left(f \frac{2\pi x}L \right).
\ee
\begin{figure}
\Fig{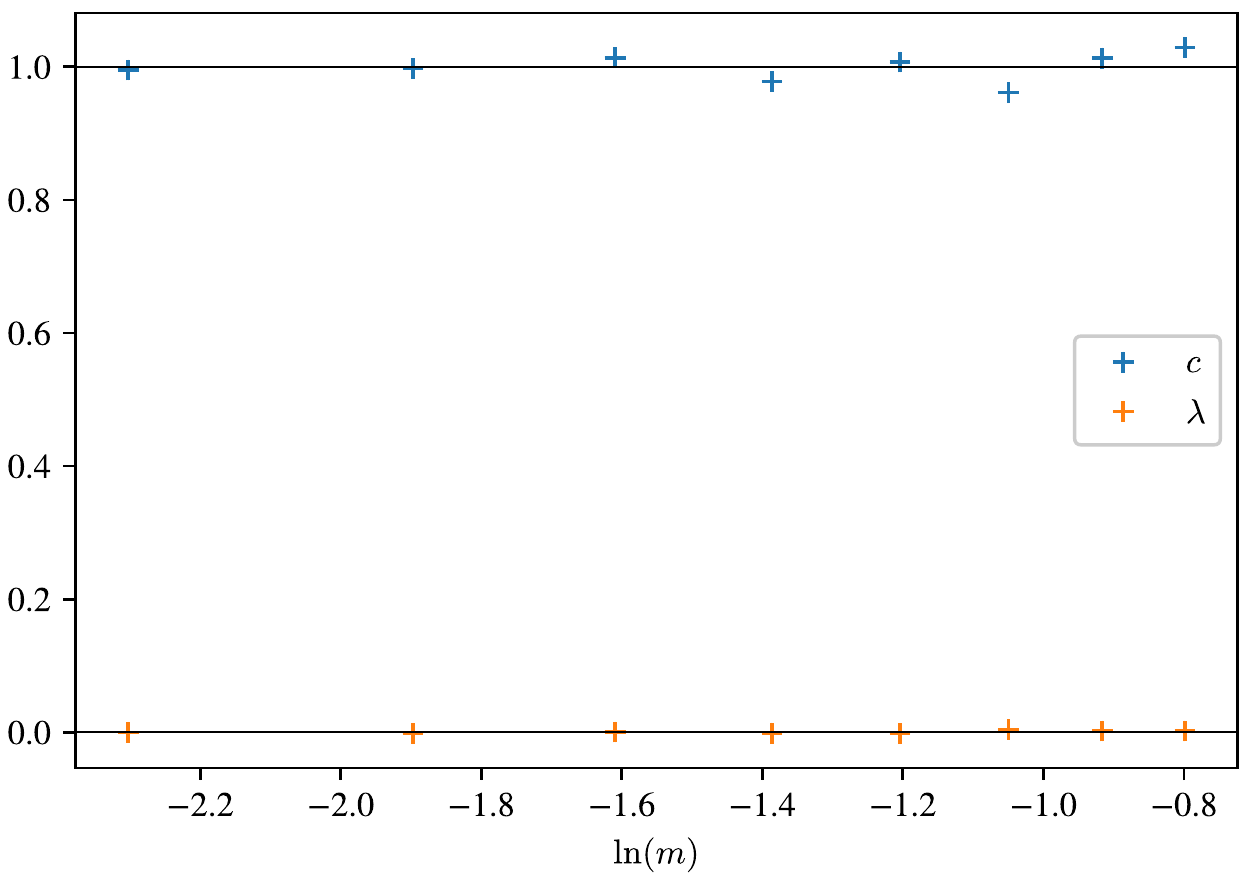}\caption{The  numerically estimated effective $c$ and $\lambda$ for the qEW equation.  The effective elasticity  $c$  does not depend on $m$ (with noticeable simulation errors at large $m$), as predicted by the statistical tilt symmetry (STS). The  numerically estimated  non-linearity $\lambda$ vanishes.}
\label{fig:qEW_algo}
\end{figure}After each avalanche, we  increase $w(x)$ by  $\delta w$ (a constant),  $w(x) \rightarrow w(x) + \delta w $.
We focus on the slowest mode $f=1$.
We then  measure the mean  interface profile, i.e.\ its {\em response}, $\overline {u(x)}$.  Varying the amplitude $A$ of the driving, we fit this response with a polynomial in $A$.
The effective parameters are then linked to the projections on these modes. 
\begin{figure}
\centering
\includegraphics[width= 0.6\linewidth]{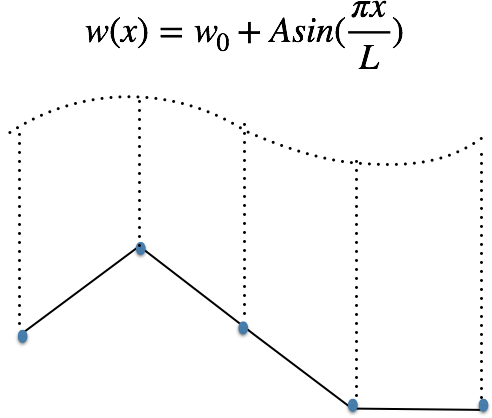}
\caption{We drive the interface with the spatially modulated driving given in \Eq{w(x)}, with $f=1$. The continuous black line and the blue dots represent the interface, while the grey dashed line represents $w(x)$.}
\label{fig:sinus_driving}
\end{figure}
To be specific, write, with $\ell:= L/2\pi$,
\bea\label{u-ansatz}
\overline {u(x) } &=& u_0(A) + u_1(A) \sin \left( \frac{x}{\ell} \right) + u_2(A) \cos\left(\frac{2x}{\ell}\right) +... \nn\\
\qquad  \\
u_0(A)& =& {^0u_0}  + {^2u_0A^2} +  \ca O(A^4),   \\
u_1(A)& =&  {^1u_1}A   + \ca O(A^3) , \\
u_2(A)& =&  {^2u_2}A^2 +   \ca O(A^4).
\label{eq:modesdvp}\eea
The dots  represent higher-order terms in $A$, while  the double-indexed $u$'s are numbers to be   estimated numerically. The lower index represents the mode, while the upper index is the order  in $A$. We inject this development into  the noiseless KPZ equation
\be
-m^2 u + c \nabla^2u +  {\lambda} (\nabla u)^2 = -m^2 A\sin\! \left( \frac{x}{\ell} \right).
\ee
It is the  non-linear term in this equation that generates the higher harmonics. The parity of the number of derivatives restricts the allowed modes to those   in \Eq{u-ansatz}. 
Matching coefficients, we find 
\bea
^0u_0 &=& w_0 \\
 {^1u_1} &=& \frac{m^2}{m^2 + \frac{c}{\ell^2}}, \\
 {^2u_0} &=& \frac{m^2 \lambda}{4\ell^2 (m^2 + \frac{c}{\ell^2})^2}, \\
  {^2u_2} &=& \frac{m^4 \lambda}{4\ell^2(m^2 + \frac{4c}{\ell^2}) (m^2 + \frac{c}{\ell^2})^2}.
\eea
These relations are   inverted to obtain $\lambda$ and $c$, 
\bea
c (m)&= &m^2 \ell^2  \frac{1- {{}^1u_1}} {^1u_1}, \\ 
 \lambda (m)&= &4 m^2\ell^2    \frac{{^2u_0}} {(^1u_1)^2}.
 \label{eq:lambda_c2}
\eea

\subsection{Tests and results}
Let us start with some tests  of our procedure for qEW. There   $\lambda(m)\equiv 0$, and  there is no renormalization of $c$, as it is  protected by the statistical-tilt symmetry, the statistical invariance of the equation of motion under the transformation $u(x,t) \rightarrow u(x,t) + \alpha x $. In Fig.~\ref{fig:qEW_algo} we show simulations for harmonic depinning (\Eq{eq:anhdep} with $c_4 =0$ and $c_2 = 1$).
We    estimated     the effective elastic constant $c$, and see that it does not renormalize and stays at $ c = 1$. Moreover,  the   numerically estimated    $\lambda=0$.

We next apply our procedure to TL92 and anharmonic depinning in $d=1$,
see Fig.~\ref{fig:fixedpoint}. For each $m$, the polynomials were fitted on $100$ different values for $A$, and each value of $A$ corresponds to a simulation of $10^5$ independent samples. The size varies from $L=512$ to $L=2048$, since for larger values of $m$ smaller systems are sufficient. 
 We find 
\bea
\psi_{c}^{d=1} &=& 1.31  \pm  0.04, \\
 \psi_{\lambda}^{d=1} &=& 0.28  \pm 0.03,
\eea in agreement with their expressions  in \Eqs{eq:scalingc2}-\eq{eq:scalinglamb},  and the numerical values given in table \ref{d=1-num-table}.

  We checked that higher-order relations (given in Appendix \ref{sec:detalgo}) give the same values for $c$ and $\lambda$. 
We further checked that the results given for $\lambda$ are the same as those obtained as a response to a tilt. (Note that to introduce a tilt with our driving protocol, one has to tilt both the driving potential and the interface.)

\begin{figure}
\mbox{\!\!\!\!\includegraphics[width=1.01\linewidth]{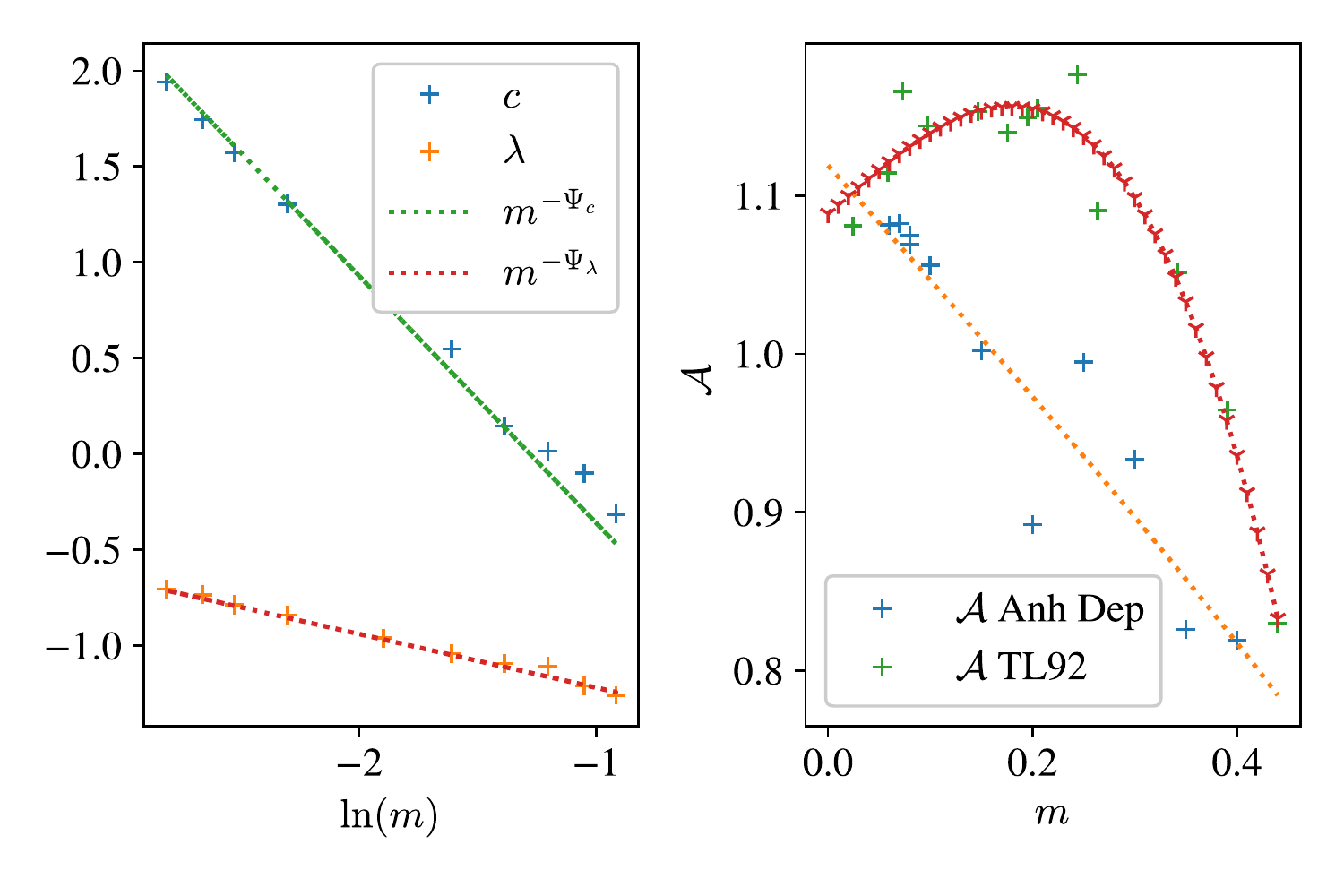}}
\caption{Left:  double-logarithmic plot for scaling of $c$  and $\lambda$ for anharmonic depinning in 1$d$   as a function of $m$. Right: Measured amplitude ratios  $\mathcal{A}$  for  TL92  and anharmonic depinning (linear plot). The dotted lines are guides for the eye. The second-order polynomial fits  show convergence to $\mathcal{A}\approx 1.10(2)$ for  $m\to 0$.}
\label{fig:fixedpoint}
\end{figure}

\begin{figure*}[t]
\centerline{\includegraphics[width=\textwidth]{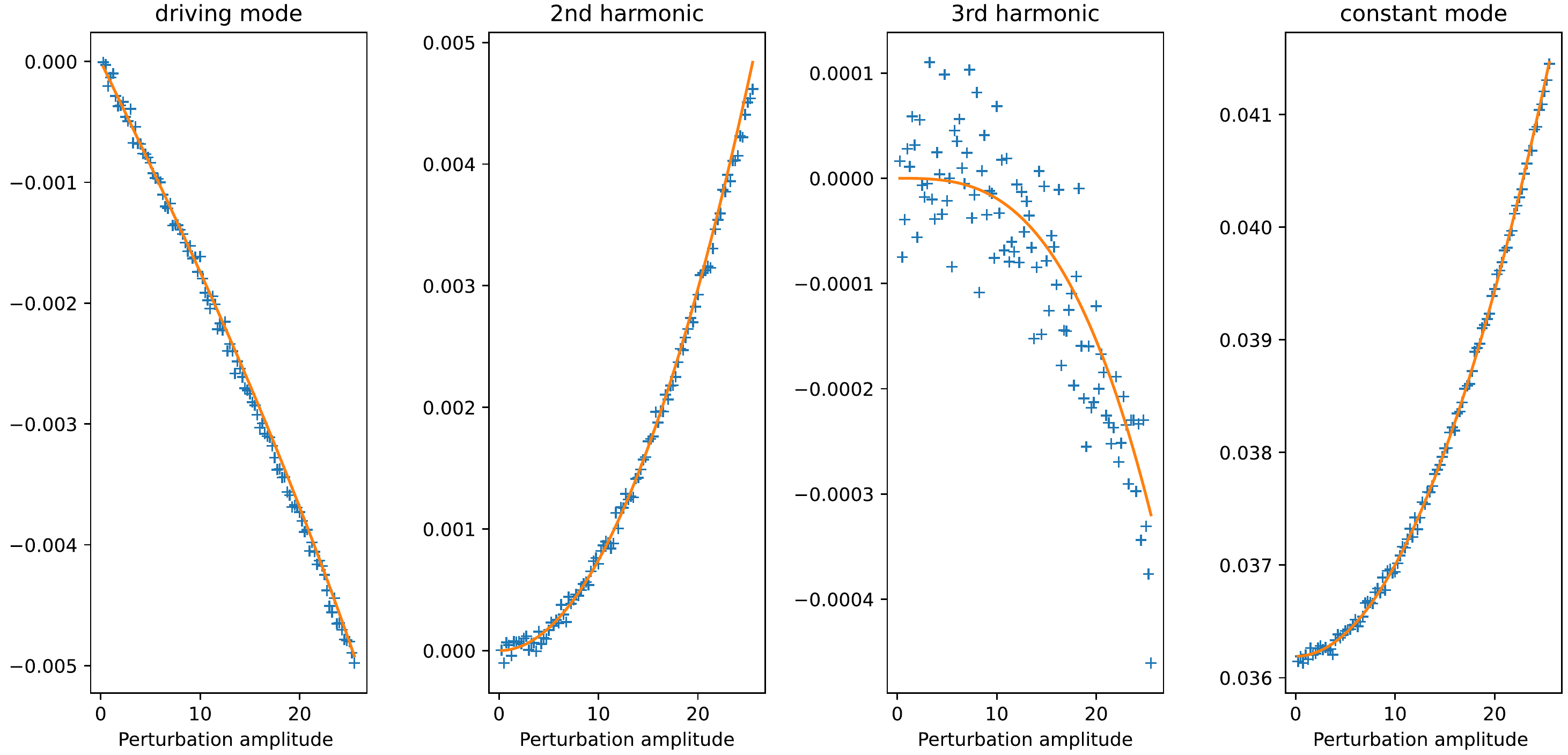}}
\caption{Numerical estimation of  the interface modes  for different driving amplitude $A$. From the orange fits  one obtains the  coefficients  $^{i}u_j$ defined in \Eqs{u-ansatz}-\eq{eq:modesdvp}. For example, from the fit of the driving mode one obtains $^1u_1$ and $^3u_1$.   The polynomial behaviour predicted in \Eq{eq:modesdvp} is   verified. The shown plots are for anharmonic depinning,   $L=1024$, and $m=0.06$. The perturbation amplitude is in lattice units.  }
\label{fig:algofits}
\end{figure*}

The determination of the effective parameters $\lambda$ and $c$ is not the only application of this algorithm: one can  numerically estimate   the effective decay of   subleading parameters present in the microscopic model, such as $c_4$, 
 and obtain valuable information on the crossover to the qKPZ universality class. This may be helpful for experiments and is summarized in Appendix \ref{sec:detalgo}. While many things can be numerically estimated, 
this technique is  limited by the available computer resources, as illustrated on Fig.~\ref{fig:c_4_decay} for the decay of $c_4$.

\subsection{The universal KPZ amplitude $\ca A$}
\label{The universal KPZ amplitude A} 
An important question is whether qKPZ is the proper large-distance description of TL92, anharmonic depinning, and itself (i.e.\ a numerical implementation of the qKPZ equation). To ensure this, the properly renormalized non-linearity $\lambda$ needs to flow to a fixed point. While $\lambda$ and $c$ both flow, i.e.\ do not go to a fixed point by themselves,  this is achieved by   the {\em universal KPZ amplitude}  $\ca A$,    
\be
 \mathcal{A}:= \rho \frac{\lambda}{c}, \qquad \rho = \frac{\Delta(0)}{| \Delta^{\prime}(0^+) |} .
 \label{eq:fixedpointA}
 \ee
The idea behind this construction is that if both $\lambda$ and $c$ are relevant, then 
\be
\lambda  \left[ \nabla u(x,t) \right] ^2 \sim c \nabla^2 u(x,t) \quad \Longrightarrow\quad \frac{\lambda }c \sim \frac 1 u
\ee
On the other hand $\Delta(u) \sim u  \Delta'(u)$, thus we can define a correlation length $\rho \sim u$ by $\rho:=\Delta(0)/|\Delta'(0^+)|$; this allows one to write the dimensionless quantity $\ca A$ in \Eq{eq:fixedpointA}. 
Note that the definition \eq{eq:fixedpointA} ensures that $\ca A$ remains invariant under a change of   units for $u$, say from mm to km, and the same (independently)  for $x$. 

The reader may wonder whether our definition for $\ca A$ is unique? It is not, as  one could instead of $\rho$  use another characteristic scale, such as   $\xi_\perp$. 
The reason we use  $\rho$ defined in \Eq{eq:fixedpointA} rather than $\xi_\perp$ defined in \Eq{xi-perp-def} is that the former is simpler to handle analytically. 

 If the qKPZ equation is the effective field theory in the limit of $m\to 0$, then the ratio $\ca A$ needs to converge to a universal limit set by the qKPZ field theory. That this is indeed the case can be seen on Fig.~\ref{fig:fixedpoint}.
In the two models simulated, the amplitude ratio converges to the same value, 
\be\label{90}
\ca A^{d=1} = 1.10(2).
\ee
Given that the microscopic models are quite different, this is a strong sign of universality.

\subsection{Interpretation of $\mathcal{A}$: How strong is the KPZ non-linearity?} \label{sec:interpet_a}
The reader may ask himself whether the amplitude $\ca A=1.1$ estimated numerically in \Eq{90} is small, or large: firstly,  
it is definitely much larger than for   qEW, for which $\ca A^{\rm qEW}=0$ (since $\lambda=0$ there).

For a rough estimate at the microscopic scale, we can  convert the arguments of section \ref{sec:mapqkpzTL92} into a prediction for $\ca A$. 
Let us first consider the mapping of TL92 onto qKPZ, which resulted into the values for $c$ and $\lambda$ given in  \Eq{23}.
If we take these values, and the lattice size $\rho=1$ for the  correlation length of the disorder, then we get
a microscopic or {\em bare} value of $\ca A$,
\be\label{A-bound-micro}
\ca A^{\rm bare}_{d=1} \approx \frac{5}4.
\ee
It is  
surprising that the estimate \eq{A-bound-micro} at small scales is close to the large-scale estimate  \eq{90} of our numerical simulation.

Can one give a bound for $\ca A$, or could one have estimated an arbitrarily large value in \Eq{90}? 
Let us consider the  drawing  of Fig.~\ref{fig:configqKPZTL92}. We ask that in absence of disorder the point $N$ does not advance, thus $F_N$ be negative. 
This means that   the ratio between KPZ\footnote{Note that    the discretization \eq{eq:discreteqkpz} for the KPZ-term induces a numerical factor of $1/4$ into the equation. Given that these values for $\lambda$ and $c$   are effective large-scale estimates, this factor should be taken with a grain of salt.}   and elastic term  is bounded by 1, 
\be
1 \ge \frac{\lambda (\nabla u)^2}{c |\nabla^2 u|} \simeq \frac{\lambda \big(\frac{\delta u}{2\xi_m}\big)^2 }{c \big(\frac {\delta u}{\xi_m}\big)^2}  = \frac{\lambda\delta u}{ 4c }  .
\ee
We  now need to estimate $\delta u$. Taking it as the typical fluctuation at scale $\xi_m$, i.e.\ the perpendicular correlation length $\xi_\bot$, gives
$\delta u=\xi_\bot$
as defined in section \ref{s:Definition of the 2-point function}.
If this heuristic argument is correct, then 
\be
\ca A \lesssim \ca A^{\rm c}:=\frac{4\rho}{\xi_\perp}.
\ee
While $\ca A^{\rm c}$ is a bound on the  value $\ca A$ can take before   the interface becomes unstable, it is not necessarily the most stringent bound.  
In our simulations we find 
\be
\frac{\rho}{\xi_\perp} = 0.85(1) , 
\ee
rather independent of $m$. 
This in turn gives  
\be
\ca A^{\rm c}_{d=1} = 3.40 (4).
\ee
Thus in $d=1$ the amplitude $\ca A$ is definitely large, though below its  critical value.  
Field theory (see the companion paper \cite{MukerjeeWiese2022}) gives a bound of  $\ca A^{\rm c}_{d=1} = 2$ (at leading order).

More intuition can be gotten from rescaling: If one uses the dimensionless variables $ \tilde u := \frac u \rho$, $ \tilde w := \frac w \rho $  and $\tilde x := \frac {x m } {\sqrt c}$,  then     blocking configurations satisfy 
\be
0 = \nabla^2 \tilde u + \mathcal{A} (\nabla \tilde u)^2 +\tilde w -\tilde u + \tilde F(\tilde x, \tilde u).
\label{eq:rescaled_qkpz}
\ee
In these units, 
forces are correlated according to 
\be \langle \tilde F(\tilde x, \tilde u)  \tilde F(\tilde x^\prime , \tilde u^\prime) \rangle  = \delta^d(\tilde x -\tilde x^\prime) \tilde{\Delta}(\tilde u - \tilde u^\prime),
\ee  with 
\be\label{100}
\tilde \Delta(\tilde u ) = \frac{m^{d-4}}{c^{d/2} \rho^2}\Delta(\rho \tilde u).
\ee 
We estimate \Eq{100} in Fig.~\ref{fig:a_delta} for $d=1$, and find that $\tilde \Delta(0)  \approx \mathcal{A} $.
\begin{figure}
\mbox{\!\!\!\!\includegraphics[width=1.0\linewidth]{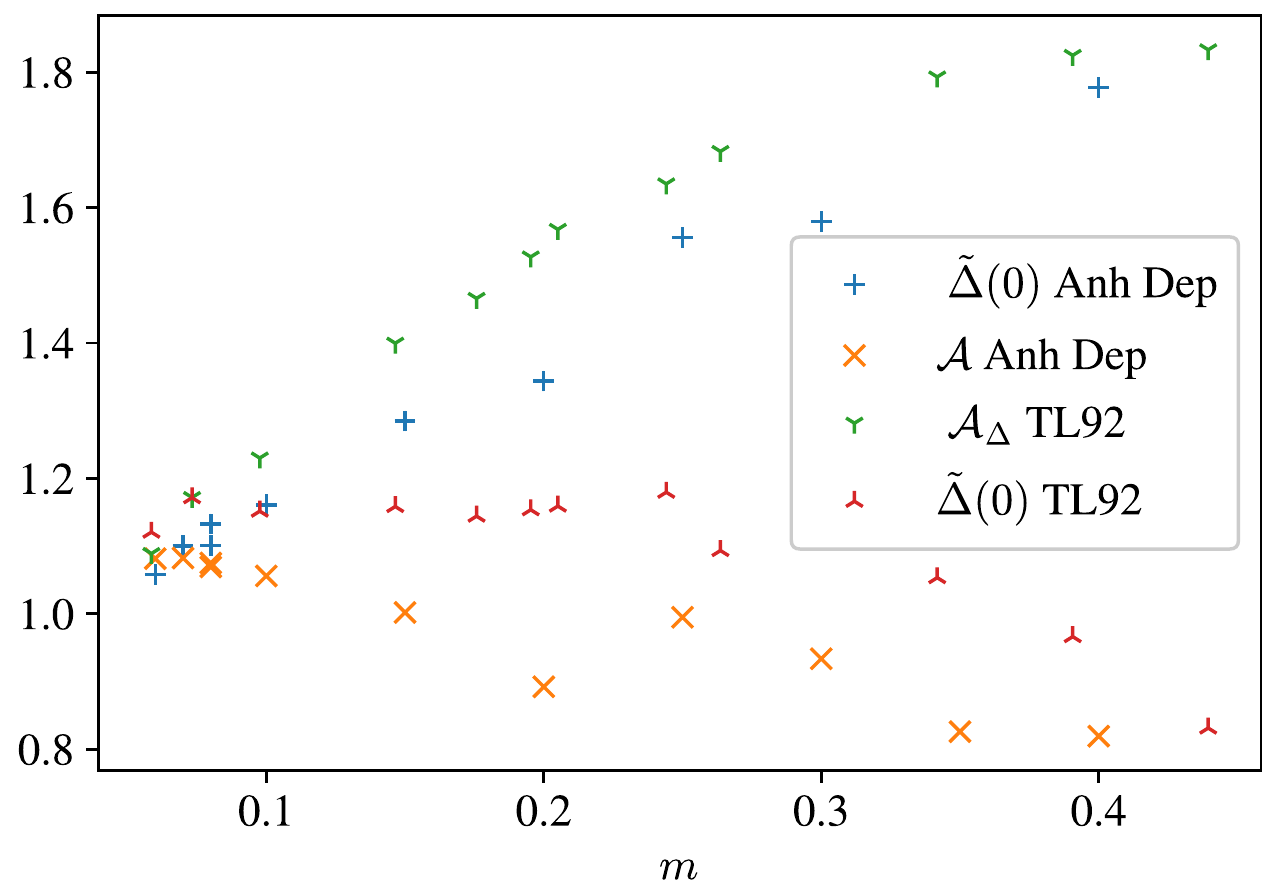}}
\caption{${\tilde \Delta(0)}$ and $ \mathcal{A}$  for Anharmonic depinning and TL92 in $d=1$. They seem to converge to the same value, for this particular scaling choice.}
\label{fig:a_delta}
\end{figure}
The conclusion is that at the  critical point, at least in $d=1$, all parameters are of order one, thus equally important.

\section{Conclusion}
\label{s:Conclusion}

We showed through theoretical arguments and numerical tests  that anharmonic depinning, qKPZ, and the cellular automaton TL92 are in the same universality class, the qKPZ universality class, for $d \leq 2$. For $2 < d \leq 4$,  there is evidence that TL92  may depart from the qKPZ universality class  (which still includes anharmonic depinning at those dimensions). 

We then elucidated the scaling relations  for  driving through a parabolic confining potential. This allowed us to understand statics and dynamics of  qKPZ. Finally, we developed an algorithm to    estimate     the renormalized (effective) coefficients of the continuity equation.  We find that, at least in $d=1$, all quantities are equally important, of order one in a particular scheme.
Our work will be used to constrain, and ultimately construct the field theory, which is  presented in a sequel to this work \cite{MukerjeeWiese2022}. 

We believe that our  technique to extract the effective coupling constants by measuring the static response of the system under spatially modulated perturbations may yield important information in other systems that lack a proper field theoretic description. As an example, we started to extend  our approach to  the thermal KPZ equation.

\begin{figure}
\centerline{\!\!\!\includegraphics[width=1.07\columnwidth]{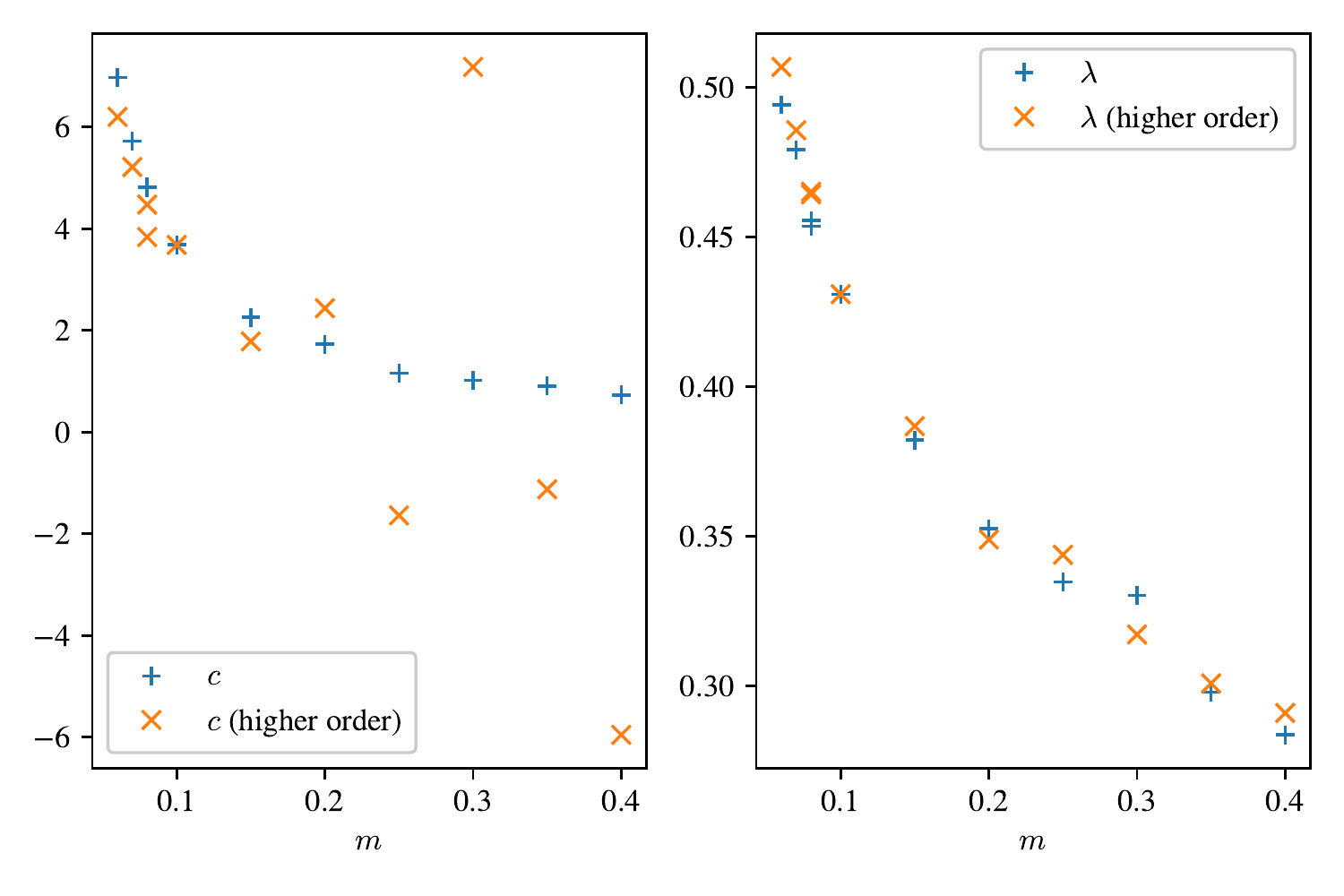}}
\caption{Comparison of the different formulae (\Eq{eq:lambda_c2} and \Eq{eq:higher_order} for a fit based on     higher-order harmonics)  for determining $c$ and $\lambda$, for aDep in $d=1$. We   see a good agreement for $\lambda$, while  for $c$ the higher-order harmonics are too noisy for $m\geq 0.2$.}
\label{fig:higherorder_c2_lamb}
\end{figure}

\acknowledgements
We thank Alberto Rosso for useful discussions. 
JAB acknowledges support from NSF grant DMS-2052616. MAM acknowledges financial support from  the Spanish Ministry and Agencia
Estatal de investigaci{\'o}n (AEI) through Project of I+D+i
Ref. PID2020-113681GB-I00, financed by MICIN/AEI/10.13039/501100011033
and FEDER “A way to make Europe”.

\appendix
 
\section{Why $z \neq 1$} \label{s:zconjecture}

In Ref.~\cite{AmaralBarabasiBuldyrevHarringtonHavlinSadr-LahijanyStanley1995} the authors provide a heuristic argument for $z=1$ in $d=1$, while  for higher dimension they conjecture that $z = d_{\min}$ with $d_{\min}$ the exponent on how the shortest-path length on a $d$ dimensional critical percolating cluster scales with the Euclidean distance. Our simulations invalidate this heuristics. Here we give  theoretical arguments as to why this heuristics fails.
Ref.~\cite{AmaralBarabasiBuldyrevHarringtonHavlinSadr-LahijanyStanley1995}  studies TL92 with parallel updates. After an avalanche, they define the path of invaded cells as the path (in $1d$) from the cell from which it was invaded to the cell it invaded. Then they assert that the path length from the start of the avalanche to site $i$  is equal to the time it took for cell $i$ to be invaded. Then $z$ is defined by $T \sim \ell^z$ with $T$ the avalanche duration and $\ell$ the lateral extension of the avalanche. Since $\xi_{\bot}/ \xi_m \rightarrow 0$,   the invading path is considered   rough as the path along the blocking configuration, so $\ell \approx \xi_m $. As a result they find $z = 1$. However, equating the length of the invading path with the duration of an avalanche is problematic: After reaching site $i$, the avalanche can   change direction and then come back, and as a result the duration is under estimated, and $z>1$.

In higher dimensions, this under-estimation persists, but is associated with another problem, that over-estimates $z$: since $\xi_{\bot}/ \xi_m \rightarrow 0$ they model the $d+1$ dimensional space in which the invading path lives as a $d$ dimensional critical percolation cluster, and then declare the invading path to be the shortest distance between two points on this percolation cluster. However, the transport properties of percolation clusters are highly dependent on the proportion of singly connected cells \cite{StaufferAharony1994} (i.e.\ cells that if removed separate the percolation cluster in two). The existence of another dimension through which the path can go  changes the statistics of those singly-connected cells. There are far more ways to reach one target, and as a result the time it takes to reach it may be smaller and $z$   over-estimated.

\begin{figure}
{\includegraphics[width=\columnwidth]{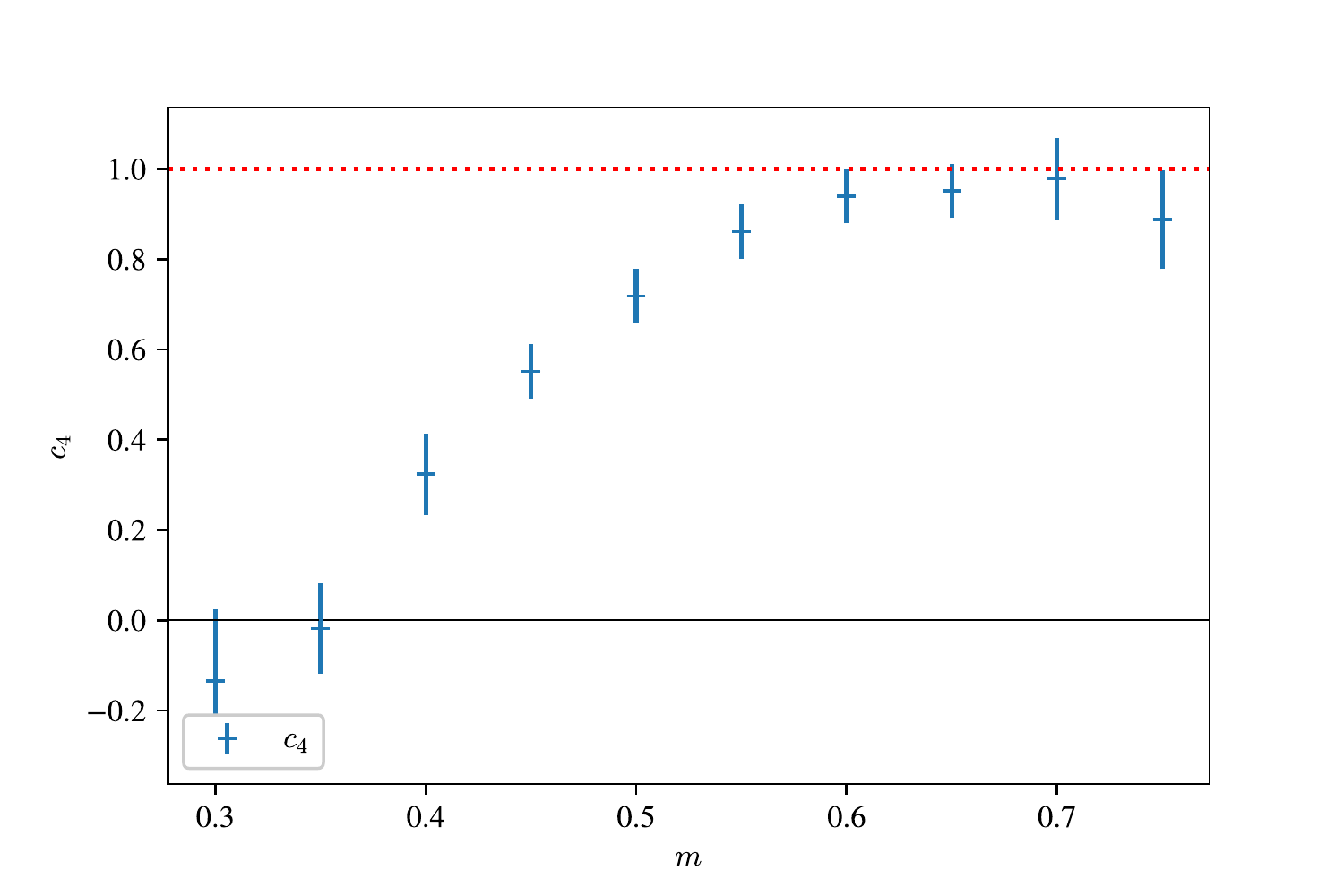}}
\caption{Numerically estimated effective anharmonicity $c_4$ determined for anharmonic depinning using   \Eq{eq:c_4}. $L=64$ and the initial condition is  $c_4 = 1$ (red dotted line). The   numerically estimated   effective $c_4 $ decreases as $m$ decreases, vanishing within the precision of our simulation when   $m \approx 0.3$. 
There  the error for $c_4$ increases, as    the small system  sees more system-spanning avalanches, crossing over to a single-particle behavior. Increasing $L$ further increases this error, since $c_4$ is determined through a high-order harmonics of the driving signal, and scales  $\sim   L^4$ appearing in \Eq{eq:c_4}. Still, qualitatively this confirms that $c_4$ can be dropped  in the effective long-distance description of the model.}
\label{fig:c_4_decay}
\end{figure}

\section{Details of the algorithm}\label{sec:detalgo}
\subsection{Numerical details}
In Fig.~\ref{fig:algofits}  we show the results of measuring the modes of the interface for different amplitudes of the driving.
One has to be careful to be in the small-perturbation limit. We find that taking the maximum perturbation amplitude to be $A=\frac L {40}$ to be appropriate. The number of points needed within that range to have a good precision on the polynomial fit is hard to deduce in advance, and varies with $m$. We find that sometimes the small perturbation limit is reached before $A=\frac L {40}$  and in that case it is good to have more points in order to maintain a good fit for the polynomials. A good rule of thumb is to have around $50$ points.

\subsection{Higher-order relations}
There are higher order relations for $\lambda$ and $c$. Here we put them, for completeness.
\bea
c(m) &=& \left(\frac{^2u_0}{^2u_2}-1\right)\frac{m^2\ell^2}{4}, 
\\
\lambda(m) &=& \frac{4\ell^2}{m^4}\left(m^2 + \frac{4c}{\ell^2}\right) \left(m^2 + \frac{c}{\ell^2}\right)^2 {^2u_2}.
\label{eq:higher_order}
\eea
In 
Fig.~\ref{fig:higherorder_c2_lamb} we can see that for smaller $m$, the higher order formulas agrees with their lower order counterpart. For higher $m$ the signal for $c$ determined with \Eq{eq:higher_order} is too noisy. For $\lambda$ there is good agreement for all $m$. 

\subsection{Crossover and higher-order anharmonic terms}
An interesting question is the crossover from the microscopic model, e.g. in anharmonic depinning which contains a coefficient $c_4$. 
How does this terms decrease with $m$? 
To answer these questions,  we derive a formula for the expression of $c_4$.
If a $c_4$ term is present, then the lowest order in $A$   is $A^3$. We find that there is a contribution of $c_4$ on the first mode, written for compactness in terms of ${}^ju_i$, $\lambda$ and $c$,
 \be
 c_4(m) =  -4\frac{\ell^4}{(^1u_1)^3} \left[\frac{ \lambda  {^1u_1} {^2u_2}}{ \ell^2} +{^3u_1 }\left(m^2  +\frac{c}{\ell^2}\right)\right].
 \label{eq:c_4}
 \ee
 $c_4$ appears as a third-order   perturbation in $A$. Since it comes from higher harmonics, it is more heavily suppressed as the system becomes larger. As a result, small system sizes (and large $m$) must be considered to   accurately estimate    $c_4$. However, there is a tradeoff, since $c_2$ and $\lambda$ are determined with a lesser accuracy for smaller systems size.
 The result for an initial anharmonic depinning equation with $c_4 = 1$ are presented in Fig.~\ref{fig:c_4_decay}. We see that at large $m$ the microscopic value $c_4=1$ is obtained. Reducing $m$ to about $0.35$, the effective $c_4$ becomes too small to be distinguishable  from the noise. 
 
%

\ifx\doi\undefined
\providecommand{\doi}[2]{\href{http://dx.doi.org/#1}{#2}}
\else
\renewcommand{\doi}[2]{\href{http://dx.doi.org/#1}{#2}}
\fi
\providecommand{\link}[2]{\href{#1}{#2}}
\providecommand{\arxiv}[1]{\href{http://arxiv.org/abs/#1}{#1}}
\providecommand{\hal}[1]{\href{https://hal.archives-ouvertes.fr/hal-#1}{hal-#1}}
\providecommand{\mrnumber}[1]{\href{https://mathscinet.ams.org/mathscinet/search/publdoc.html?pg1=MR&s1=#1&loc=fromreflist}{MR#1}}

\tableofcontents

\vfill

\end{document}